\documentclass[twocolumn,showpacs,%
  aps,superscriptaddress,%
  eqsecnum,prd,notitlepage,showkeys,10pt]{revtex4-1}

\usepackage{amssymb}
\usepackage{amsmath}
\usepackage{graphicx}
\usepackage{dcolumn}
\usepackage{hyperref}
\usepackage[labelfont=bf,justification=raggedright,singlelinecheck=false]{caption, subcaption}
\usepackage{color}

\begin{document}
\title{Liquid xenon scintillation measurements and pulse shape discrimination in the LUX dark matter detector}


\author{D.S.~Akerib} \affiliation{Case Western Reserve University, Department of Physics, 10900 Euclid Ave, Cleveland, OH 44106, USA} \affiliation{SLAC National Accelerator Laboratory, 2575 Sand Hill Road, Menlo Park, CA 94205, USA} \affiliation{Kavli Institute for Particle Astrophysics and Cosmology, Stanford University, 452 Lomita Mall, Stanford, CA 94309, USA}
\author{S.~Alsum} \affiliation{University of Wisconsin-Madison, Department of Physics, 1150 University Ave., Madison, WI 53706, USA}  
\author{H.M.~Ara\'{u}jo} \affiliation{Imperial College London, High Energy Physics, Blackett Laboratory, London SW7 2BZ, United Kingdom}  
\author{X.~Bai} \affiliation{South Dakota School of Mines and Technology, 501 East St Joseph St., Rapid City, SD 57701, USA}  
\author{A.J.~Bailey} \affiliation{Imperial College London, High Energy Physics, Blackett Laboratory, London SW7 2BZ, United Kingdom}  
\author{J.~Balajthy} \affiliation{University of Maryland, Department of Physics, College Park, MD 20742, USA}  
\author{P.~Beltrame} \affiliation{SUPA, School of Physics and Astronomy, University of Edinburgh, Edinburgh EH9 3FD, United Kingdom}  
\author{E.P.~Bernard} \affiliation{University of California Berkeley, Department of Physics, Berkeley, CA 94720, USA} \affiliation{Yale University, Department of Physics, 217 Prospect St., New Haven, CT 06511, USA} 
\author{A.~Bernstein} \affiliation{Lawrence Livermore National Laboratory, 7000 East Ave., Livermore, CA 94551, USA}  
\author{T.P.~Biesiadzinski} \affiliation{Case Western Reserve University, Department of Physics, 10900 Euclid Ave, Cleveland, OH 44106, USA} \affiliation{SLAC National Accelerator Laboratory, 2575 Sand Hill Road, Menlo Park, CA 94205, USA} \affiliation{Kavli Institute for Particle Astrophysics and Cosmology, Stanford University, 452 Lomita Mall, Stanford, CA 94309, USA}
\author{E.M.~Boulton} \affiliation{University of California Berkeley, Department of Physics, Berkeley, CA 94720, USA} \affiliation{Lawrence Berkeley National Laboratory, 1 Cyclotron Rd., Berkeley, CA 94720, USA} \affiliation{Yale University, Department of Physics, 217 Prospect St., New Haven, CT 06511, USA}
\author{P.~Br\'as} \affiliation{LIP-Coimbra, Department of Physics, University of Coimbra, Rua Larga, 3004-516 Coimbra, Portugal}  
\author{D.~Byram} \affiliation{University of South Dakota, Department of Physics, 414E Clark St., Vermillion, SD 57069, USA} \affiliation{South Dakota Science and Technology Authority, Sanford Underground Research Facility, Lead, SD 57754, USA} 
\author{M.C.~Carmona-Benitez} \affiliation{Pennsylvania State University, Department of Physics, 104 Davey Lab, University Park, PA 16802-6300, USA} \affiliation{University of California Santa Barbara, Department of Physics, Santa Barbara, CA 93106, USA} 
\author{C.~Chan} \affiliation{Brown University, Department of Physics, 182 Hope St., Providence, RI 02912, USA}  
\author{A.~Currie} \affiliation{Imperial College London, High Energy Physics, Blackett Laboratory, London SW7 2BZ, United Kingdom}  
\author{J.E.~Cutter} \affiliation{University of California Davis, Department of Physics, One Shields Ave., Davis, CA 95616, USA}  
\author{T.J.R.~Davison} \affiliation{SUPA, School of Physics and Astronomy, University of Edinburgh, Edinburgh EH9 3FD, United Kingdom}  
\author{A.~Dobi} \affiliation{Lawrence Berkeley National Laboratory, 1 Cyclotron Rd., Berkeley, CA 94720, USA}  
\author{E.~Druszkiewicz} \affiliation{University of Rochester, Department of Physics and Astronomy, Rochester, NY 14627, USA}  
\author{B.N.~Edwards} \affiliation{Yale University, Department of Physics, 217 Prospect St., New Haven, CT 06511, USA}  
\author{S.R.~Fallon} \affiliation{University at Albany, State University of New York, Department of Physics, 1400 Washington Ave., Albany, NY 12222, USA}  
\author{A.~Fan} \affiliation{SLAC National Accelerator Laboratory, 2575 Sand Hill Road, Menlo Park, CA 94205, USA} \affiliation{Kavli Institute for Particle Astrophysics and Cosmology, Stanford University, 452 Lomita Mall, Stanford, CA 94309, USA} 
\author{S.~Fiorucci} \affiliation{Lawrence Berkeley National Laboratory, 1 Cyclotron Rd., Berkeley, CA 94720, USA} \affiliation{Brown University, Department of Physics, 182 Hope St., Providence, RI 02912, USA} 
\author{R.J.~Gaitskell} \affiliation{Brown University, Department of Physics, 182 Hope St., Providence, RI 02912, USA}  
\author{J.~Genovesi} \affiliation{University at Albany, State University of New York, Department of Physics, 1400 Washington Ave., Albany, NY 12222, USA}  
\author{C.~Ghag} \affiliation{Department of Physics and Astronomy, University College London, Gower Street, London WC1E 6BT, United Kingdom}  
\author{M.G.D.~Gilchriese} \affiliation{Lawrence Berkeley National Laboratory, 1 Cyclotron Rd., Berkeley, CA 94720, USA}  
\author{C.R.~Hall} \affiliation{University of Maryland, Department of Physics, College Park, MD 20742, USA}  
\author{S.J.~Haselschwardt} \affiliation{University of California Santa Barbara, Department of Physics, Santa Barbara, CA 93106, USA}  
\author{S.A.~Hertel} \affiliation{University of Massachusetts, Amherst Center for Fundamental Interactions and Department of Physics, Amherst, MA 01003-9337 USA} \affiliation{Lawrence Berkeley National Laboratory, 1 Cyclotron Rd., Berkeley, CA 94720, USA} \affiliation{Yale University, Department of Physics, 217 Prospect St., New Haven, CT 06511, USA}
\author{D.P.~Hogan} \affiliation{University of California Berkeley, Department of Physics, Berkeley, CA 94720, USA}  
\author{M.~Horn} \affiliation{South Dakota Science and Technology Authority, Sanford Underground Research Facility, Lead, SD 57754, USA} \affiliation{University of California Berkeley, Department of Physics, Berkeley, CA 94720, USA} \affiliation{Yale University, Department of Physics, 217 Prospect St., New Haven, CT 06511, USA}
\author{D.Q.~Huang} \affiliation{Brown University, Department of Physics, 182 Hope St., Providence, RI 02912, USA}  
\author{C.M.~Ignarra} \affiliation{SLAC National Accelerator Laboratory, 2575 Sand Hill Road, Menlo Park, CA 94205, USA} \affiliation{Kavli Institute for Particle Astrophysics and Cosmology, Stanford University, 452 Lomita Mall, Stanford, CA 94309, USA} 
\author{R.G.~Jacobsen} \affiliation{University of California Berkeley, Department of Physics, Berkeley, CA 94720, USA}  
\author{W.~Ji} \affiliation{Case Western Reserve University, Department of Physics, 10900 Euclid Ave, Cleveland, OH 44106, USA} \affiliation{SLAC National Accelerator Laboratory, 2575 Sand Hill Road, Menlo Park, CA 94205, USA} \affiliation{Kavli Institute for Particle Astrophysics and Cosmology, Stanford University, 452 Lomita Mall, Stanford, CA 94309, USA}
\author{K.~Kamdin} \affiliation{University of California Berkeley, Department of Physics, Berkeley, CA 94720, USA}  
\author{K.~Kazkaz} \affiliation{Lawrence Livermore National Laboratory, 7000 East Ave., Livermore, CA 94551, USA}  
\author{D.~Khaitan}\email{dkhaitan@u.rochester.edu} \affiliation{University of Rochester, Department of Physics and Astronomy, Rochester, NY 14627, USA}  
\author{R.~Knoche} \affiliation{University of Maryland, Department of Physics, College Park, MD 20742, USA}  
\author{B.G.~Lenardo}\email{bglenardo@ucdavis.edu} \affiliation{University of California Davis, Department of Physics, One Shields Ave., Davis, CA 95616, USA} \affiliation{Lawrence Livermore National Laboratory, 7000 East Ave., Livermore, CA 94551, USA}  \affiliation{Stanford University, Department of Physics, 382 Via Pueblo, Stanford, CA 94305, USA}
\author{K.T.~Lesko} \affiliation{Lawrence Berkeley National Laboratory, 1 Cyclotron Rd., Berkeley, CA 94720, USA}  
\author{J.~Liao} \affiliation{Brown University, Department of Physics, 182 Hope St., Providence, RI 02912, USA}  
\author{A.~Lindote} \affiliation{LIP-Coimbra, Department of Physics, University of Coimbra, Rua Larga, 3004-516 Coimbra, Portugal}  
\author{M.I.~Lopes} \affiliation{LIP-Coimbra, Department of Physics, University of Coimbra, Rua Larga, 3004-516 Coimbra, Portugal}  
\author{A.~Manalaysay} \affiliation{University of California Davis, Department of Physics, One Shields Ave., Davis, CA 95616, USA}  
\author{R.L.~Mannino} \affiliation{Texas A \& M University, Department of Physics, College Station, TX 77843, USA} \affiliation{University of Wisconsin-Madison, Department of Physics, 1150 University Ave., Madison, WI 53706, USA} 
\author{M.F.~Marzioni} \affiliation{SUPA, School of Physics and Astronomy, University of Edinburgh, Edinburgh EH9 3FD, United Kingdom}  
\author{D.N.~McKinsey} \affiliation{University of California Berkeley, Department of Physics, Berkeley, CA 94720, USA} \affiliation{Lawrence Berkeley National Laboratory, 1 Cyclotron Rd., Berkeley, CA 94720, USA} \affiliation{Yale University, Department of Physics, 217 Prospect St., New Haven, CT 06511, USA}
\author{D.-M.~Mei} \affiliation{University of South Dakota, Department of Physics, 414E Clark St., Vermillion, SD 57069, USA}  
\author{J.~Mock} \affiliation{University at Albany, State University of New York, Department of Physics, 1400 Washington Ave., Albany, NY 12222, USA}  
\author{M.~Moongweluwan} \affiliation{University of Rochester, Department of Physics and Astronomy, Rochester, NY 14627, USA}  
\author{J.A.~Morad} \affiliation{University of California Davis, Department of Physics, One Shields Ave., Davis, CA 95616, USA}  
\author{A.St.J.~Murphy} \affiliation{SUPA, School of Physics and Astronomy, University of Edinburgh, Edinburgh EH9 3FD, United Kingdom}  
\author{C.~Nehrkorn} \affiliation{University of California Santa Barbara, Department of Physics, Santa Barbara, CA 93106, USA}  
\author{H.N.~Nelson} \affiliation{University of California Santa Barbara, Department of Physics, Santa Barbara, CA 93106, USA}  
\author{F.~Neves} \affiliation{LIP-Coimbra, Department of Physics, University of Coimbra, Rua Larga, 3004-516 Coimbra, Portugal}  
\author{K.~O'Sullivan} \affiliation{University of California Berkeley, Department of Physics, Berkeley, CA 94720, USA} \affiliation{Lawrence Berkeley National Laboratory, 1 Cyclotron Rd., Berkeley, CA 94720, USA} \affiliation{Yale University, Department of Physics, 217 Prospect St., New Haven, CT 06511, USA}
\author{K.C.~Oliver-Mallory} \affiliation{University of California Berkeley, Department of Physics, Berkeley, CA 94720, USA}  
\author{K.J.~Palladino} \affiliation{University of Wisconsin-Madison, Department of Physics, 1150 University Ave., Madison, WI 53706, USA} \affiliation{SLAC National Accelerator Laboratory, 2575 Sand Hill Road, Menlo Park, CA 94205, USA} \affiliation{Kavli Institute for Particle Astrophysics and Cosmology, Stanford University, 452 Lomita Mall, Stanford, CA 94309, USA}
\author{E.K.~Pease} \affiliation{University of California Berkeley, Department of Physics, Berkeley, CA 94720, USA} \affiliation{Lawrence Berkeley National Laboratory, 1 Cyclotron Rd., Berkeley, CA 94720, USA} \affiliation{Yale University, Department of Physics, 217 Prospect St., New Haven, CT 06511, USA}
\author{C.~Rhyne} \affiliation{Brown University, Department of Physics, 182 Hope St., Providence, RI 02912, USA}  
\author{S.~Shaw} \affiliation{University of California Santa Barbara, Department of Physics, Santa Barbara, CA 93106, USA} \affiliation{Department of Physics and Astronomy, University College London, Gower Street, London WC1E 6BT, United Kingdom} 
\author{T.A.~Shutt} \affiliation{Case Western Reserve University, Department of Physics, 10900 Euclid Ave, Cleveland, OH 44106, USA}  \affiliation{Kavli Institute for Particle Astrophysics and Cosmology, Stanford University, 452 Lomita Mall, Stanford, CA 94309, USA}
\author{C.~Silva} \affiliation{LIP-Coimbra, Department of Physics, University of Coimbra, Rua Larga, 3004-516 Coimbra, Portugal}  
\author{M.~Solmaz} \affiliation{University of California Santa Barbara, Department of Physics, Santa Barbara, CA 93106, USA}  
\author{V.N.~Solovov} \affiliation{LIP-Coimbra, Department of Physics, University of Coimbra, Rua Larga, 3004-516 Coimbra, Portugal}  
\author{P.~Sorensen} \affiliation{Lawrence Berkeley National Laboratory, 1 Cyclotron Rd., Berkeley, CA 94720, USA}  
\author{T.J.~Sumner} \affiliation{Imperial College London, High Energy Physics, Blackett Laboratory, London SW7 2BZ, United Kingdom}  
\author{M.~Szydagis} \affiliation{University at Albany, State University of New York, Department of Physics, 1400 Washington Ave., Albany, NY 12222, USA}  
\author{D.J.~Taylor} \affiliation{South Dakota Science and Technology Authority, Sanford Underground Research Facility, Lead, SD 57754, USA}  
\author{W.C.~Taylor} \affiliation{Brown University, Department of Physics, 182 Hope St., Providence, RI 02912, USA}  
\author{B.P.~Tennyson} \affiliation{Yale University, Department of Physics, 217 Prospect St., New Haven, CT 06511, USA}  
\author{P.A.~Terman} \affiliation{Texas A \& M University, Department of Physics, College Station, TX 77843, USA}  
\author{D.R.~Tiedt} \affiliation{South Dakota School of Mines and Technology, 501 East St Joseph St., Rapid City, SD 57701, USA}  
\author{W.H.~To} \affiliation{California State University Stanislaus, Department of Physics, 1 University Circle, Turlock, CA 95382, USA} \affiliation{SLAC National Accelerator Laboratory, 2575 Sand Hill Road, Menlo Park, CA 94205, USA} \affiliation{Kavli Institute for Particle Astrophysics and Cosmology, Stanford University, 452 Lomita Mall, Stanford, CA 94309, USA}
\author{M.~Tripathi} \affiliation{University of California Davis, Department of Physics, One Shields Ave., Davis, CA 95616, USA}  
\author{L.~Tvrznikova} \affiliation{University of California Berkeley, Department of Physics, Berkeley, CA 94720, USA} \affiliation{Lawrence Berkeley National Laboratory, 1 Cyclotron Rd., Berkeley, CA 94720, USA} \affiliation{Yale University, Department of Physics, 217 Prospect St., New Haven, CT 06511, USA}
\author{U.~Utku} \affiliation{Department of Physics and Astronomy, University College London, Gower Street, London WC1E 6BT, United Kingdom}  
\author{S.~Uvarov} \affiliation{University of California Davis, Department of Physics, One Shields Ave., Davis, CA 95616, USA}  
\author{V.~Velan} \affiliation{University of California Berkeley, Department of Physics, Berkeley, CA 94720, USA}  
\author{J.R.~Verbus} \affiliation{Brown University, Department of Physics, 182 Hope St., Providence, RI 02912, USA}  
\author{R.C.~Webb} \affiliation{Texas A \& M University, Department of Physics, College Station, TX 77843, USA}  
\author{J.T.~White} \affiliation{Texas A \& M University, Department of Physics, College Station, TX 77843, USA}  
\author{T.J.~Whitis} \affiliation{Case Western Reserve University, Department of Physics, 10900 Euclid Ave, Cleveland, OH 44106, USA} \affiliation{SLAC National Accelerator Laboratory, 2575 Sand Hill Road, Menlo Park, CA 94205, USA} \affiliation{Kavli Institute for Particle Astrophysics and Cosmology, Stanford University, 452 Lomita Mall, Stanford, CA 94309, USA}
\author{M.S.~Witherell} \affiliation{Lawrence Berkeley National Laboratory, 1 Cyclotron Rd., Berkeley, CA 94720, USA}  
\author{F.L.H.~Wolfs} \affiliation{University of Rochester, Department of Physics and Astronomy, Rochester, NY 14627, USA}  
\author{J.~Xu} \affiliation{Lawrence Livermore National Laboratory, 7000 East Ave., Livermore, CA 94551, USA}  
\author{K.~Yazdani} \affiliation{Imperial College London, High Energy Physics, Blackett Laboratory, London SW7 2BZ, United Kingdom}  
\author{S.K.~Young} \affiliation{University at Albany, State University of New York, Department of Physics, 1400 Washington Ave., Albany, NY 12222, USA}  
\author{C.~Zhang} \affiliation{University of South Dakota, Department of Physics, 414E Clark St., Vermillion, SD 57069, USA}  
\collaboration{LUX Collaboration}



\begin{abstract}
Weakly Interacting Massive Particles (WIMPs) are a leading candidate for dark matter and are expected to produce nuclear recoil (NR) events within liquid xenon time-projection chambers. We present a measurement of the scintillation timing characteristics of liquid xenon in the LUX dark matter detector and develop a pulse shape discriminant to be used for particle identification. To accurately measure the timing characteristics, we develop a template-fitting method to reconstruct the detection times of photons. Analyzing calibration data collected during the 2013-16 LUX WIMP search, we provide a new measurement of the singlet-to-triplet scintillation ratio for electron recoils (ER) below 46~keV, and we make a first-ever measurement of the NR singlet-to-triplet ratio at recoil energies below 74~keV. We exploit the difference of the photon time spectra for NR and ER events by using a prompt fraction discrimination parameter, which is optimized using calibration data to have the least number of ER events that occur in a 50\% NR acceptance region. We then demonstrate how this discriminant can be used in conjunction with the charge-to-light discrimination to possibly improve the signal-to-noise ratio for nuclear recoils.
\end{abstract}

\maketitle
\section{Introduction}
\label{sec:intro}

Liquid xenon time projection chamber (TPC) experiments are leaders in sensitivity to the interactions of Weakly Interacting Massive Particles (WIMPs), a class of as-yet-unobserved particles that have been proposed as a solution to the dark matter problem \cite{LUXRun4PRL,PandaXSpinIndependent2016, XENON100Result}. In such experiments, the WIMP is predicted to scatter elastically from a xenon nucleus, resulting in a nuclear recoil (NR). The primary backgrounds are electron recoils (ER) from gamma and beta radiation released by residual radioactivity in the detector materials, with a small contribution from neutrino-electron scattering. Interactions in liquid xenon produce scintillation photons and ionization electrons which can be measured to reconstruct information about the interaction. TPC experiments measure both the ionization and scintillation signals and use this information to reconstruct the energy deposition, particle type, and the position of the interaction.

Background rejection is paramount to the success of liquid xenon dark matter searches. Material screening and shielding are the primary methods to mitigate backgrounds; detectors are constructed from highly radiopure materials and are operated in well-shielded underground environments to reduce backgrounds from cosmic rays and environmental sources. Position reconstruction allows fiducialization and the rejection of multiple-scattering events. The former eliminates ER backgrounds from detector materials stopping close the edges of the sensitive volume, while the latter removes event topologies inconsistent with WIMP scattering. Background events which remain in the data can be rejected through particle-type discrimination between ER and NR ~\cite{LUXRun4PRL}. In liquid xenon TPC experiments, this last step is typically done using the ratio of ionization charge to scintillation light in the event, which is higher for ER events than NR events. The present work explores enhancing the ER background rejection using pulse shape discrimination (PSD) applied to the scintillation signal alone.

Scintillation light is produced by the self-trapping of excited xenon atoms (Xe$^*$), created when a particle deposits energy in the liquid. Direct excitation and recombination of electron--ion pairs create excited atoms, which combine with a neutral ground-state Xe atom to form the molecular dimer Xe$^*_2$. The dimer decays to the monatomic ground state via emission of a VUV photon ($\lambda$ = 175~nm) \cite{MartinExcimers,FUJII2015293}. These two processes are shown schematically in Eq.~\ref{diag:directscint} (direct excitation) and Eq.~\ref{diag:recombscint} (recombination of electron--ion pair).
\begin{equation}
\label{diag:directscint}
\begin{split}
\text{Xe}^* + \text{Xe} & \,\,\rightarrow\,\, \text{Xe}_2^* \\
& \,\,\rightarrow\,\, \text{Xe} + \text{Xe} + \gamma\, ,
\end{split}
\end{equation}
\begin{equation}
\label{diag:recombscint}
\begin{split}
\text{Xe}^+ + \text{Xe} & \,\,\rightarrow\,\, \text{Xe}^+_2 \\
\text{Xe}^+_2 + e^- & \,\,\rightarrow \,\, \text{Xe}^{**} + \text{Xe} \\
\text{Xe}^{**} & \,\,\rightarrow\,\,\text{Xe}^* + \text{heat} \\
\text{Xe}^* + \text{Xe} & \,\,\rightarrow\,\, \text{Xe}^*_2 \\
& \,\,\rightarrow\,\, \text{Xe} + \text{Xe} + \gamma\, .
\end{split}
\end{equation}
The decay of the dimer is observed to have both a fast and a slow component, which are interpreted as de-excitation of the singlet $^1\Sigma^+_u$ and the triplet $^3\Sigma^+_u$ states, respectively \cite{KubotaTriplet,Hitachi1983PulseShape}. There are conflicting measurements of the lifetimes of these states in the literature; measurements of the singlet time constant $\tau_1$ range from 2 to 4~ns, while measurements of the triplet time constant $\tau_3$ range from 21 to 28~ns \cite{KubotaRecombination,Hitachi1983PulseShape}. Both components have been observed for electron recoils, alpha recoils, and recoiling fission fragments. 

For electron recoils, some experiments operating without an applied electric field have observed a time profile that is best fit with a single exponential with $\tau$~=~30--45~ns \cite{Dawson2005,Akimov2002,Hitachi1983PulseShape}. This is attributed to an additional time delay due to electron--ion recombination. This interpretation is supported by measurements that show that the scintillation time structure reduces to the characteristic singlet/triplet shape under an applied electric field (which suppresses recombination) \cite{KubotaTriplet}. In addition, recent measurements, without an applied field, show an energy-dependence of the long component, correlated with the energy-dependence of recombination~\cite{XMASSPulseShape2016}. No field-dependence is observed for alpha particle or fission fragment recoils, suggesting that recombination-related timing effects are only significant at low ionization densities. At the energies (0-50~keV) and electric fields (100--1000~V/cm) relevant for modern liquid xenon TPC experiments, there are no direct measurements of the effects of recombination on ER scintillation timing. However, extrapolating to this regime using the empirical model given in Ref.~\cite{MockNESTPulseShapes} suggests that recombination may not play a significant role in scintillation emission timing in these experiments, and that pulse shapes can be well-described purely in terms of the singlet and triplet emission.

The ratio of singlet emission to triplet emission varies with particle type, opening up the possibility for ER/NR discrimination using PSD. Multiple groups have studied liquid xenon PSD in small R\&D detectors \cite{Akimov2002,Kwong2010,Ueshima2011}, and it was successfully used to reduce backgrounds in early liquid-xenon-based dark matter searches \cite{DAMAXe,ZEPLIN_I,XENON10InelasticDM,XMASSLightWIMPSearch}. However, these studies are restricted to small detectors or detectors with spherical photosensor coverage of the xenon volume. Current and future TPC experiments have meter-scale dimensions and make extensive use of reflectors to maximize light collection \cite{XENON1T_PhysicsReach, LZ_CDR}. In such detectors, scintillation pulse shapes are subject to significant distortion from scattering, reflection, and absorption of photons by detector materials. In addition, previous studies have not attempted to reconstruct the singlet/triplet ratio for both ER and NR pulses at the low energies relevant to dark matter searches. Attempts to simulate scintillation pulses must therefore rely on measurements at higher energies, which may not accurately reflect xenon microphysics in the region of interest.

In this work, we present a measurement of scintillation characteristics and PSD in the LUX detector, a $\sim$0.5~m$\times$0.5~m cylindrical liquid xenon TPC~\cite{Akerib20121}. We study both ER and NR calibration data taken throughout the LUX WIMP-search campaign. First, a template-based photon reconstruction algorithm is used to deconvolve the response of the electronics and photosensors in order to reconstruct the time when a photon strikes a photomultiplier tube. The spectra of photon detection times are added across many pulses to construct average pulse shapes for both ER and NR events. Second, we develop an analytical model to decouple detector effects from xenon scintillation emission. This model is fit to data to extract physical parameters that can inform simulation packages such as NEST \cite{NESTpaper}. Finally, we construct a pulse shape discriminant using the prompt-fraction technique, and compute the power of PSD background rejection in LUX. Using the best-fit parameters from the analytical model, we construct a simulation that accurately reproduces PSD measurements from data. The discrimination power improves with recoil energy, and we demonstrate how PSD can be used in conjunction with the charge-to-light ratio to further improve background rejection. These features make it attractive for exotic dark matter searches in which low-energy recoils are suppressed, such as searches for momentum-dependent and inelastic dark matter scattering from nuclei \cite{Fitzpatrick2012_EFT,Bramante_Inelastic}. These measurements allow estimation of the PSD capabilities of the current and next generation of liquid xenon dark matter experiments, and can be applied to future dark matter searches using the LUX dataset.


\section{The LUX Experiment}\label{sec:detector}

The LUX detector is a dual-phase xenon TPC designed to detect WIMP scattering with xenon nuclei. It was operated from April~2013 through June~2016 in the Davis Cavern at the Sanford Underground Research Facility (SURF) in Lead, South Dakota \cite{SURFDocument}. Dark matter search data were acquired in two exposure periods, denoted WS2013 and WS2014-16 \cite{LUXRun4PRL}. To meet the stringent low-background requirements required for the dark matter search, the detector is located deep underground (4,300 meters water equivalent overburden), is surrounded by a 7.6~m tall by 6.1~m diameter water shield, and is constructed from materials that have been carefully screened for radiopurity. The sensitive volume is approximately 48~cm in height and 24~cm in radius, and contains $\sim$250~kg of liquid xenon. Each end of the TPC is instrumented with an array of 61 Hamamatsu R8778 photomultipler tubes (PMTs) to detect light signals generated in the TPC. Twelve PTFE panels, $>$95\% reflective at 175~nm~\cite{Neves2017}, line the walls to increase the light collection efficiency. The scintillation signal, denoted S1, is detected directly by the PMTs. Ionization electrons are drifted under an applied electric field and extracted into a gas region at the top of the detector producing an electroluminescence signal, denoted S2. The  $\left(x,y\right)$ position of the events is reconstructed using the pattern of S2 light on the top PMT array~\cite{LUXPositionReconstruction}, while the depth is reconstructed from the time delay between the S1 and S2 signals. The energy deposition of the event is reconstructed from the magnitudes of the two signals.

The PMT signals are routed to an external electronics breakout box for processing, before digitization. Signals are amplified in two stages at the pre- and post-amplifiers, which provide a total effective gain of 7.5. The signals are shaped by a 30~MHz low-pass filter. The resulting single photoelectron (SPE) pulses have a full width at half maximum (FWHM) of $>$20~ns \cite{FahamThesis}. Signals are digitized using a 100~MHz Struck SIS3301 8-channel fast analog to digital converter (ADC). The average digitized area of an SPE is $\sim$100~mV$\cdot$ns \cite{Akerib20121}. Recently, it has been shown that VUV photons have a $\sim$20\% probability of generating two photoelectrons at the photocathode of the R8778 PMT \cite{Faham_VUV_doublePE}. S1 and S2 pulse areas are therefore measured in units of detected photons (phd) rather than photoelectrons. In addition, a ``spike count" has been employed in the LUX dark matter analyses to improve resolution for very low-energy S1 pulses \cite{LUXRun3Reanalysis,LUXRun4PRL}. This method is not used in the present work, as we are focused on higher energy events.

Calibration campaigns were conducted throughout the exposure period to monitor detector stability and response. Detector stability, electron lifetime, and signal corrections were measured using a $^{83\mathrm{m}}$Kr source dissolved in the xenon~\cite{LUXKrPaper}. 
These calibrations occurred weekly throughout both exposure periods. The low-energy NR and ER responses were periodically calibrated {\it in--situ} using fast neutrons from deuterium-deuterium fusion (DD) \cite{LUXDD} and electrons from the beta decays of tritium \cite{LUXTritium}. Neutrons from the DD generator traveled through an air-filled collimating pipe suspended in the water tank, and were approximately normally incident on the detector at a level $\sim$7.5~cm below the LXe surface in the TPC. The tritium source was deployed in the form of tritiated methane (CH$_3$T) and, as with the $^{83m}$Kr, was mixed into the detector through the xenon circulation system. Both DD and CH$_3$T calibrations were performed at the end of each WIMP search run, as well as three times during WS2014-16. To calibrate the depth-dependent response to NR events, DD calibrations were conducted at different heights at the start and end of WS2014-16. Additional calibration campaigns were carried out at the end of WS2014-16, including an injection of $^{14}$C into the xenon circulation system. The higher energy beta spectrum provided by $^{14}$C (endpoint at 156~keV) provides a source of ER events beyond the 18~keV endpoint of tritium. 

We use all of the DD, CH$_3$T, and $^{14}$C calibration data in the analysis presented here. Due to limited statistics in the lower portion of the detector, we developed our analysis and fit our analytical model using the data in the top drift bin from the WS2014-16 WIMP analysis ($t_{drift} = $40--105~$\mu$s) and demonstrate consistency between data and simulations in the lower drift bins.

\section{Photon timing}
\label{sec:photontiming}
In past studies, the time structure of detected scintillation light was typically obtained by measuring the shape of pulses summed over all channels in a detector. However, the $\sim$20~ns shaping time constant and 10~ns sampling period of the LUX DAQ are similar to the timescale to the de-excitation process of liquid xenon, and may therefore mask underlying  scintillation characteristics. For this analysis, we developed a photon timing algorithm and a channel-to-channel time calibration technique that accurately reconstructs a photon's detection time by deconvolving it from the electronic pulse and correcting for relative offsets.  

\subsection{Photon timing algorithm}

Precise timing is achieved with an analysis technique that separates pulses into individual detected photons, similar to the approaches in Refs.~\cite{XMASSPulseShape2016,AkashiRonquestPSD2015}. After baseline subtraction and normalization by PMT gains, the waveforms in individual PMT channels were analyzed in three steps: 
$1)$ template model fitting, 
$2)$ template model selection, and 
$3)$ re-weighting of the reconstructed photons. 

In the template fitting stage, the waveform in a single channel is fit with five separate $n$-photon models, composed of the sum of up to five single-photon template functions (the restriction to $n \leq 5$ is expected to be more than $99\%$ efficient for scintillation pulses up to 300 detected photons). The single-photon template is an empirical model constructed from an average of 1,000 waveforms with areas between 0.5--1.5~phd. The fit is performed using the Migrad routine built into the TMinuit class in ROOT~\cite{ROOT}, with the amplitude and the arrival time of each template as free parameters. Initial values for times and amplitudes are given by the time and height of the peaks in the waveform, defined as maxima above a threshold of 0.1~phd/sample ($\sim$5$\sigma$ above baseline fluctuations). When there are fewer than five peaks, the photon fit is repeated with all possible permutations, allowing multiple detected photons piling up to form a single spike. The resulting fits must meet two criteria: none of the reconstructed photons may have an area less than 0.15~phd, and the time separation between all pairs of photons must be greater than one sample. The first criterion removes fits in which we reconstruct fluctuations in baseline noise. Roughly 2\% of real photons fall below this threshold. The second criterion removes fits with multiple photons reconstructed within a single sample, where our algorithm is unable to accurately separate photons (a correction is applied later to account for unresolved pileup). If a particular fit fails one or both of these cuts. The best-fit times, areas, and likelihood values for each of the remaining $n$-photon models are passed to the next stage of the algorithm for comparison and selection.

The model comparison stage uses Bayes' Theorem to assign a likelihood score to each $n$-photon model, and the model with the highest likelihood score is selected. The likelihood score is the product of the maximum likelihood from the fit and a prior probability calculated using the measured area. For a given waveform (denoted $D$) and $n$-photon model (denoted $\mathcal{M}_n$), Bayes' Theorem can be used to calculate the probability of $\mathcal{M}_n$ given $D$, with
\begin{equation}
P(\mathcal{M}_n|D) = \frac{P(D|\mathcal{M}_n) P(\mathcal{M}_n)}{P(D)}\, .
\end{equation}
Here, $P(D|\mathcal{M}_n)$ is the maximum likelihood given by the fit, $P(D)$ is a flat normalization constant which we ignore, and $P(\mathcal{M}_n)$ is the prior. The prior $P(\mathcal{M}_n)$ is the probability of measuring the observed area if there were actually $n$ detected photons in the channel. This is calculated using a single-photon area response PDF, averaged over all PMT channels, which incorporates the $\sim20$\% probability of xenon scintillation light producing two photoelectrons in the R8778 PMTs~\cite{Faham_VUV_doublePE}. The $P(\mathcal{M}_n)$ prior discourages overfitting by applying a penalty to models composed of many reconstructed photons with improbably small areas. The model with the largest overall likelihood score $P(\mathcal{M}_n | D)$ is selected and we return the best-fit arrival times and amplitudes.

\begin{figure}
{\includegraphics[width=0.5\textwidth]{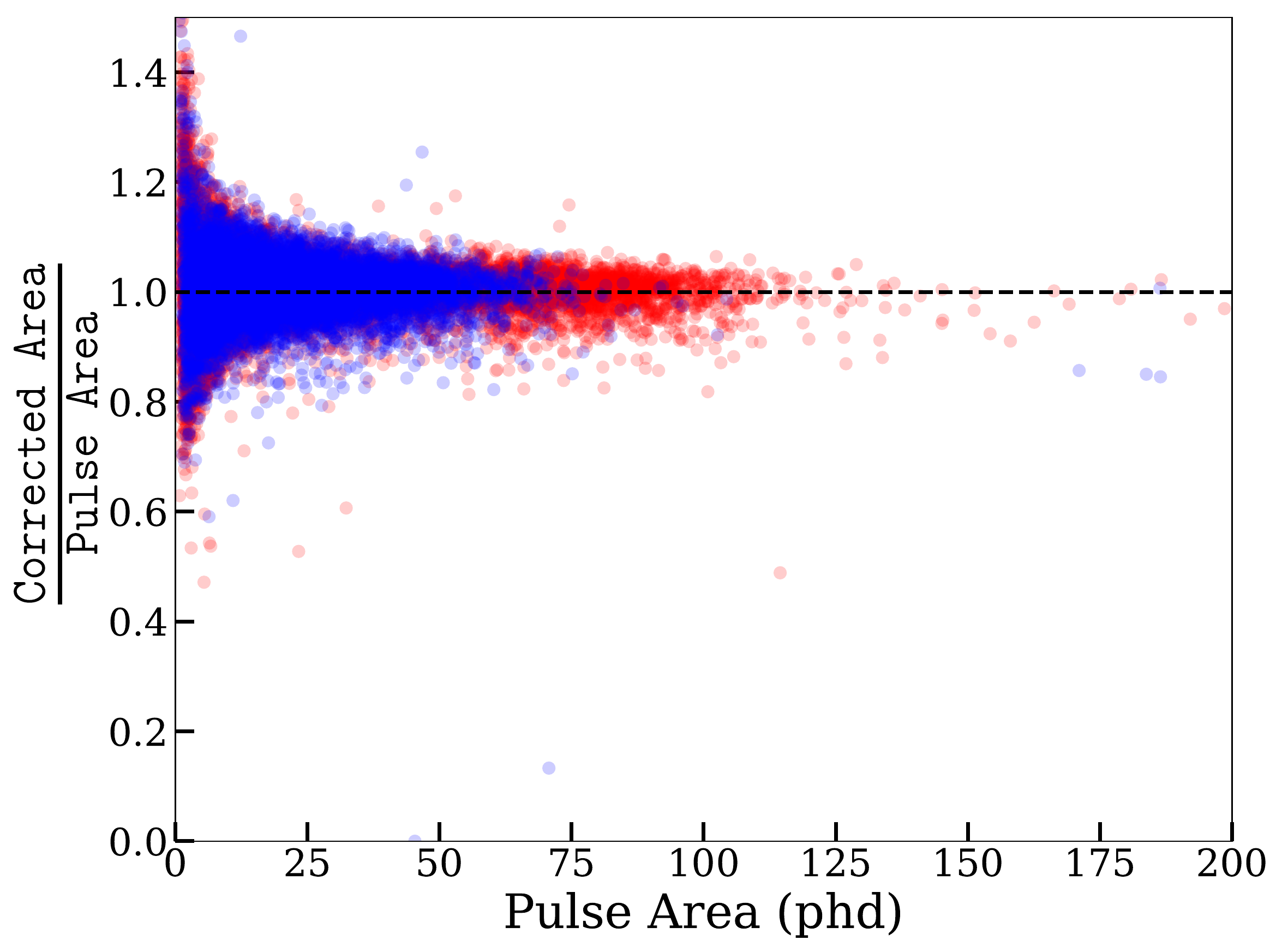}}
\caption{Ratio of corrected to raw S1 pulse area in detected photons (phd) for both tritium (blue) and DD neutron (red) calibration data. The corrected photon count is computed using Eq.~\ref{eq:corrected_count} with $C = 1.04$. The black dashed line has a slope $m = 1$, and is shown for comparison.}
\label{fig:count_vs_area}
\end{figure}

To correct for unresolved pile-up, we assign a weight to each reconstructed photon that is equal to the area of the fitted template. Pile-up occurs when the fitting algorithm fails to split a single peak, usually when two photons arrive in a single channel within 1 sample. The total number of photons counted in the pulse is given by,
\begin{equation}
\text{Corrected photon count} = C \sum_{i=1}^N w_i\, ,
\label{eq:corrected_count}
\end{equation}
where $w_i$ is the weight of the $i$-th photon, $N$ is the uncorrected number of photons returned by the fits in all channels, and $C$ is an overall correction factor. The latter accounts for inefficiencies that may arise due to the fit threshold or inexact area matching of the template function with true pulses.
To find $C$, we fit a linear model of the form $y = m\,x$ to the DD neutron data, where $x$ is the pulse area (in phd) and $y = \sum w_i$. Then $C = 1/m$. We find that $C=1.04 \pm 0.01$ reproduces the total number of photons obtained from the pulse area. A comparison between our corrected photon count and the pulse area using these values is shown in Fig.~\ref{fig:count_vs_area}. The pulse area and the corrected photon count agree throughout the 0--200~phd pulse area range used in this work.

\begin{figure}
{\includegraphics[width=0.5\textwidth]{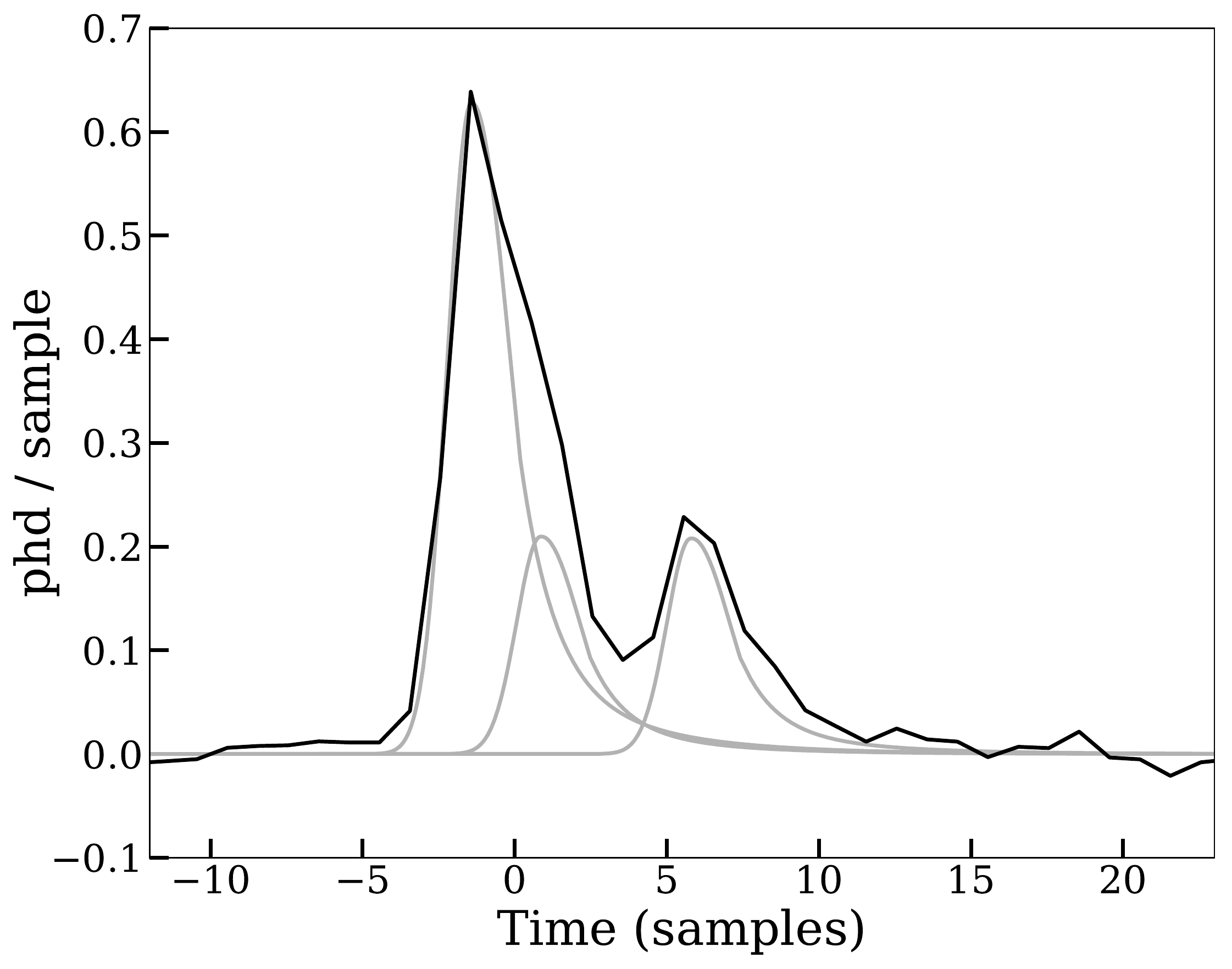}}
\caption{Example simulated waveform (black). Three photons arriving at $t$ = $-$0.75, 1.60, and 6.50 samples (1~sample = 10~ns) are used to generate the simulated signal. The photon timing algorithm described in Section~\ref{sec:photontiming} reconstructs 3 photons arriving at $t$ = $-$0.76, 1.51, and 6.45 samples (grey). }
\label{fig:sim_waveform}
\end{figure}

\begin{figure}
\subcaptionbox{Measured Scintillation Pulse}{\includegraphics[width=0.5\textwidth]{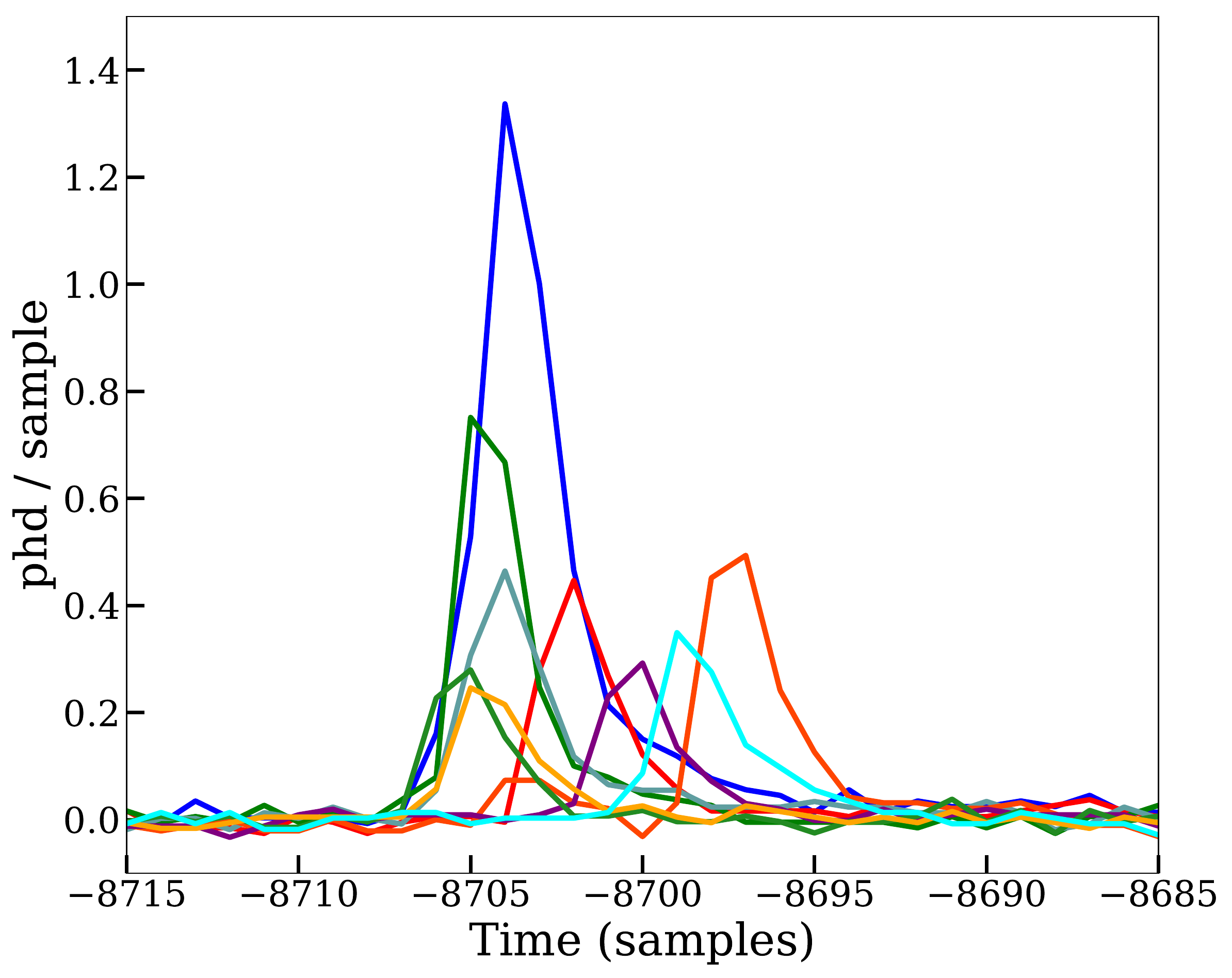}}
\subcaptionbox{Reconstructed Photon Times}{\includegraphics[width=0.5\textwidth]{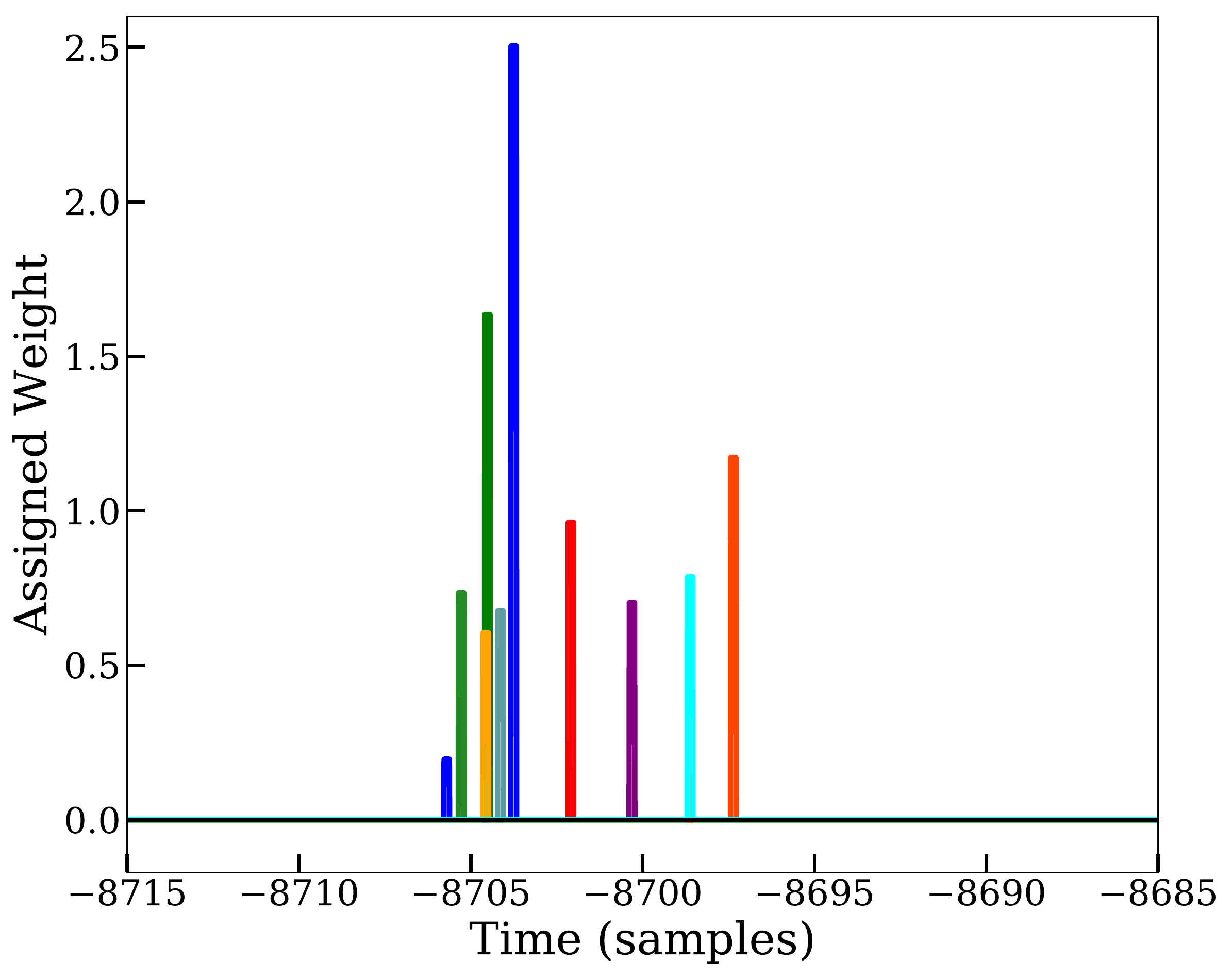}}
\caption{\emph{(a)} A scintillation pulse from a CH$_3$T calibration event, separated by PMT channel. \emph{(b)} The reconstructed peak times for the photons in the pulse is shown. The times are weighted by the fitted area of the template to correct for unresolved pile-up of multiple photons arriving in a single channel.}
\label{fig:S1_example_timing}
\end{figure}

The times of the photons returned by the fits correspond to the photon detection times, deconvolved from the shaping of the detector electronics. The algorithm is demonstrated visually in Figs.~\ref{fig:sim_waveform} and \ref{fig:S1_example_timing}. Figure~\ref{fig:sim_waveform} shows an example best-fit model with a simulated pulse. Although there are only two peaks in the pulse, the model selection algorithm correctly prefers the 3-photon model and reconstructs the times to within 0.1~samples (1~ns) for each photon. Fig.~\ref{fig:S1_example_timing} shows the algorithm applied to a real S1 pulse from the tritium calibration data. We estimate the 1$\sigma$ uncertainty of the photon detection time from the fit to be 1.6~ns, calculated from the average uncertainty returned by the Migrad fitter when our algorithm is applied to the CH$_3$T calibration data. 

\hspace{1ex}
\subsection{Channel-to-Channel Time Calibration}

\begin{figure}
{\label{fig:top_pmt_near}}{\includegraphics[width=0.5\textwidth]{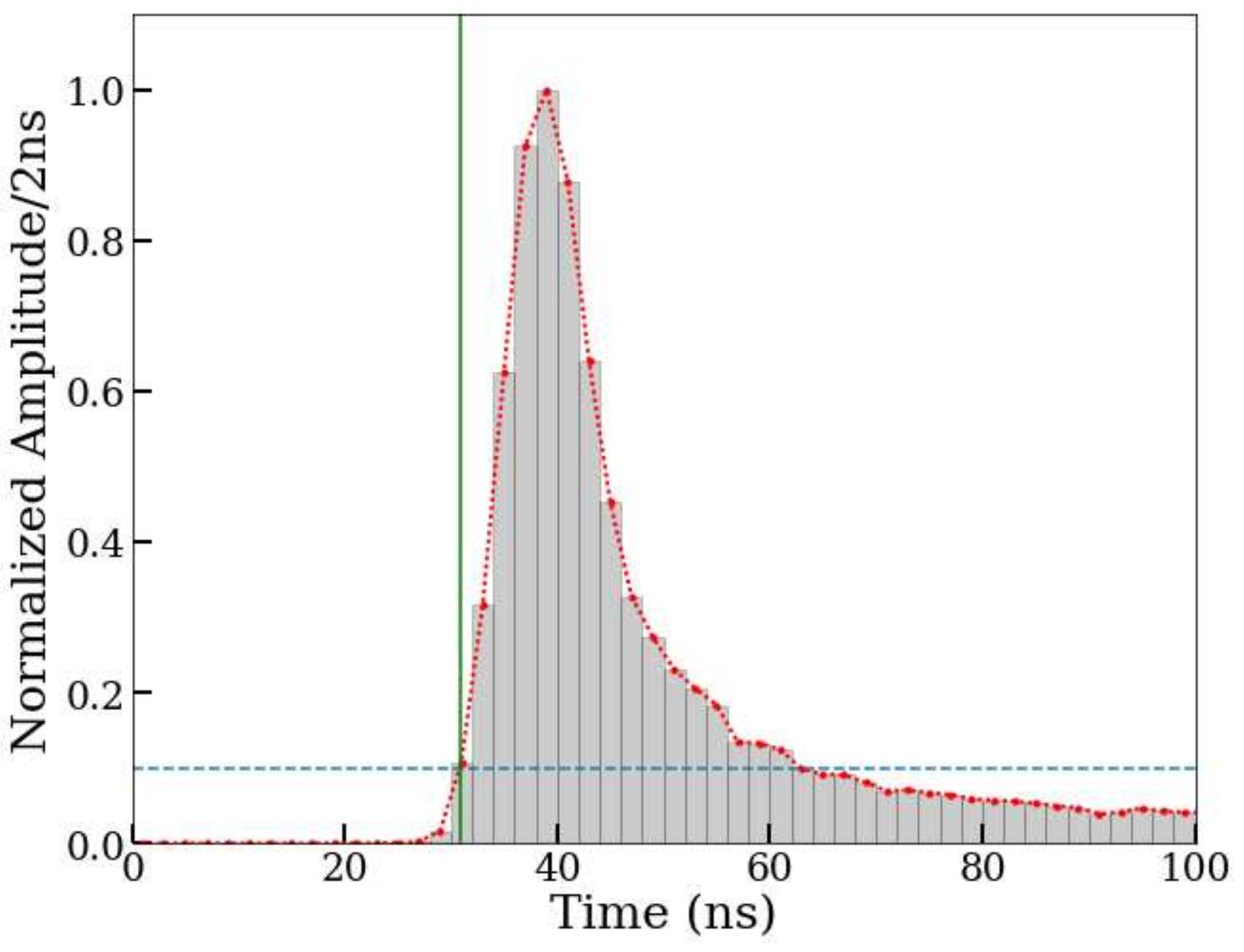}}
\caption{An example of a photon detection time distribution for a PMT located on the bottom array, facing the strobed LED on the top array. The photon detection times shown are measured relative to the LED trigger and are histogrammed in 2~ns bins. The points between the data are interpolated with a linear spline, which is shown with the dotted red trace. The dashed blue line shows 10\% of the peak height and is compared to the spline fit to obtain the offset for that channel, shown by the solid green trace.}
\label{fig:relative_timing}
\end{figure}

There are several factors that affect relative timing between PMT channels. The R8778 PMTs are specified to have an electron transit time of 41.0~$\pm$~1.7~($\sigma$)~ns at 1500~V; this transit time varies inversely with the square root of the bias voltage \cite{PMTHandbook}. Gain-matching of the PMTs in LUX requires operational voltages to vary between 1000 and 1500~V, which causes the electron transit time to vary between 41 and 50~ns. Differences in cable lengths can cause further differences in signal arrival times. The shaping filters on the pre- and post-amplifiers further degrade timing accuracy and may add relative delays between channels.  Finally, a 100~MHz clock pulse is propagated to each digitizer that can cause synchronization delays between digitizers \cite{Akerib20121}. All of these relative offsets must be measured and corrected for in this analysis so that coincident photons are correctly aligned in time.

We measure the combined effect of these time offsets using LEDs mounted on the top and bottom PMT arrays. The LED system includes twelve 440~nm diodes, capped with PTFE diffusers, that are used for gain and after-pulsing calibrations of the PMTs. To measure relative time offsets, pulses with a FWHM of 20~ns, a rise/fall-time of 5~ns, and a peak amplitude in the range 3.36--3.80~V are propagated to individual LEDs within the chamber. The resulting single photoelectron pulses in each channel are fit with the single-photon template to determine their arrival time relative to the LED strobe. The direct-path travel time for a photon from an LED to a PMT is subtracted to remove photon path length differences from this calibration. If the PMT is located on the same array as the LED, there is no direct optical path from the LED to the PMT; we therefore assume that the shortest path is via reflection off the liquid-gas interface. For each channel, a distribution of the path-corrected photon detection time, relative to the LED trigger, is obtained. A typical distribution is shown in Fig.~\ref{fig:relative_timing}.

A common reference time needs to be selected for each channel to serve as the correction to be applied to that channel. Tails longer than the strobe time of the LED are observed in these time distributions and are attributed to photons scattering within the detector volume. To avoid these effects from biasing our measurement, 10\% of the peak amplitude on the rising edge of each distribution is used to define the correction time for each channel. These corrections vary by up to 20~ns (2~samples) from channel to channel. Measurements are repeated with two LEDs in the top array and two LEDs in the bottom array to test for systematic effects from PMT coverage. The corrections between different LED measurements agree to within 2~ns, which represents the resolution of the measurement; we take this as our $1\sigma$ uncertainty. The corrections are subtracted from the reconstructed photon times in the analysis presented in Section~\ref{sec:analysis}.


\section{Liquid Xenon Scintillation in the LUX detector}
\label{sec:modeling}

\subsection{Analytical model of photon detection times}
The analytical model of photon detection time in LUX is built from three components: scintillation emission, optical transport, and a model of instrument response. 

The scintillation emission distribution is assumed to be of the form
\begin{equation}
P_s(t) = C_1\,\, e^{-t/\tau_1} + C_3\,\, e^{-t/\tau_3}\, ,
\label{eq:scint_dist}
\end{equation}
where $\tau_1$ and $\tau_3$ are the time constants governing the decay of the singlet and triplet states. In this parameterization, the ratio of singlet photons to triplet photons, referred to as the intensity ratio or the singlet/triplet ratio in the literature, is given by $ \left( C_1\,\tau_1 \right)/ \left(C_3\,\tau_3\right)$. Additional timing effects in electron recoils due to electron--ion recombination are neglected in our model as they are expected to be suppressed by the applied electric field and the high linear energy transfer (LET) at low energies. For the calibration data used to constrain the model ($\sim$300~V/cm and $\sim$10~keV), an empirical formula in Ref.~\cite{MockNESTPulseShapes} predicts a recombination time scale of $\tau_{rec} < 0.7$~ns. As this is significantly smaller than the other timescales in this analysis, we neglect a full treatment and simply use two different triplet time constant for ER and NR, $\tau_{3_{er}}$ and $\tau_{3_{nr}}$. Any recombination effects will be absorbed by $\tau_{3_{er}}$ and will result in $\tau_{3_{er}}$ slightly larger than $\tau_{3_{nr}}$.

{\setlength{\tabcolsep}{0.7em}
\begin{table}[t!]
\def\arraystretch{1.3}
\caption{Optical transport parameters at different locations inside the LUX detector (see Eq.~\ref{eq:ref_dist}). The first four rows correspond to the drift bins used in the WS2014-16 dark matter analysis \cite{LUXRun4PRL}, while the D-D beam location is used to constrain our pulse shape model in Sec.~\ref{subsec:fits}.}
\begin{center}
\begin{tabular}{ l c c c c }
Position & $A$ & $B_a$ & $\tau_a$~(ns) & $\tau_b$~(ns) \\
\hline
\hline
Top & 0.0544 & 1.059 & 11.2 & 2.80 \\
Top center & 0.0489 & 1.017 & 11.2 & 5.21 \\
Bottom center & 0.0586 & 0.906 & 11.2 & 1.56 \\ 
Bottom & 0.120 & 0.798 & 11.1 & 1.68 \\
D-D beam & 0.0574 & 1.062 & 11.1 & 2.70
\end{tabular}

\label{tab:transport}
\end{center}
\end{table}}

\begin{figure}[t!]
{\includegraphics[width=0.5\textwidth]{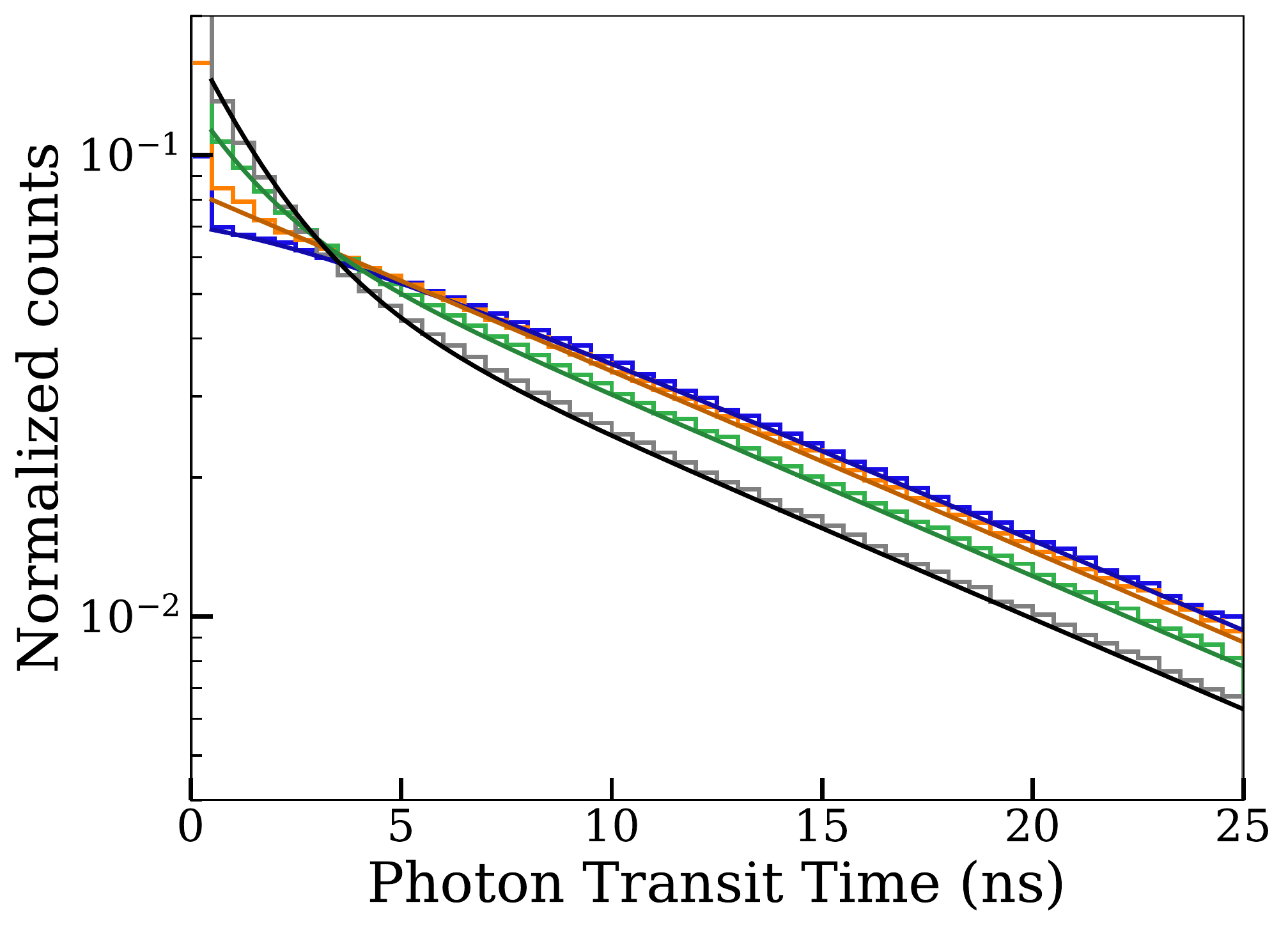}}
\caption{Simulated time distribution of photons detected in the LUX detector, calculated using the ray-tracing capabilities in the LUXSim Geant4 simulation package. The photon sources are 10~cm thick, 20~cm diameter cylinders, centered at different depths below the liquid surface: 42~cm (grey), 32~cm (green), 22~cm (orange), and 12~cm (blue). These slices in depth correspond to the drift time bins used in the WS2014-16 dark matter search \cite{LUXRun4PRL}. Solid lines show the second term in Eq.~\ref{eq:ref_dist}, fitted to these simulations. The excess in the first bin from direct-path photon arrivals is parameterized by a constant $A$ and is not reflected in the curves shown. Values of the parameters for each position are given in Table~\ref{tab:transport}
} 
\label{fig:optical}
\end{figure}

The optical transport distribution is constructed using photon-tracking simulations which take into account physical and geometrical effects on xenon scintillation light inside the LUX detector \cite{MongkolFlightPath}. Photon transport times depend on several physical properties of the detector internals: reflection and absorption at internal surfaces, reflection and transmission at the liquid/gas interface, and absorption and scattering in the liquid. The values for these parameters are constrained $in$-$situ$ using $^{83\text{m}}$Kr calibration data~\cite{LUX_PRD}. Photon transport is simulated using the LUXSim package~\cite{LUXSim_Paper}, a Geant4-based Monte Carlo code~\cite{Geant4_1,Geant4_2}. Figure~\ref{fig:optical} shows the optical transport distributions in each of four height bins used in the LUX WIMP search analysis with the WS2014-16 exposure \cite{LUXRun4PRL}. The differences between the simulations reflect the depth-dependent probability, due to the combination of geometric efficiency and reflection at the liquid-gas interface, for photons to travel directly to PMTs. To include optical transport in the analytical model, we introduce the empirical distribution
\begin{equation}
P_o(t) = A\,\delta(t) + \left( 1-A \right) \left[ \frac{B_a}{\tau_a}\,e^{-t/\tau_a} + \frac{B_b}{\tau_b}\,e^{-t/\tau_b} \right]\, ,
\label{eq:ref_dist}
\end{equation}
where $A$, $B_a$, $\tau_a$, and $\tau_b$ are fitted to the simulated distributions. The first term is a Dirac delta function which parameterizes the light which travels directly to a PMT, while the second term parameterizes the time distribution from light that reflects and scatters from the detector internals. Normalization requires $B_b~=~1-B_a$. The uncertainties in the optical parameters, given in Ref.~\cite{LUX_PRD}, are a source of systematic error in our analysis. We discuss this further in Section~\ref{subsec:fits}. 

Finally, we treat instrumental effects as normally distributed variables, parameterized by an overall width $\sigma$. There are two leading sources of random timing fluctuations in our data: the transit time spread of the R8778 PMTs $\sigma_{tts} = 1.9$~ns (at the average operating bias of $\sim$1300~V) \cite{R8778Datasheet}, and the uncertainty in the reconstructed detection time from the photon timing algorithm, $\sigma_{fit} = 1.6$~ns. There is also the 2~ns uncertainty in the channel-to-channel time corrections, $\sigma_{tc}$. While this is a fixed time offset for each channel rather than a random pulse-by-pulse fluctuation, the result is a net smearing of the pulse shape when averaging pulses together across all of the channels. Using simulations, the net effect of the time correction uncertainties on average pulse shapes was determined to be equivalent to normally-distributed random fluctuations. Therefore, the overall width added to photon time spectra by the effects in the electronics and the data reduction pipeline can be described by adding these three effects in quadrature:
\begin{equation}
\sigma = \sqrt{\sigma_{tts}^2 + \sigma_{tc}^2 + \sigma_{fit}^2}\, .
\label{eq:sigma}
\end{equation}

The total analytical model for photon detection time is given by the convolution of Eq.~\ref{eq:scint_dist}, Eq.~\ref{eq:ref_dist}, and a Gaussian distribution of width $\sigma$, given by Eq.~\ref{eq:sigma}:
\begin{equation}
\begin{split}
P(t) = \\
\sum_{i=1,3}\, \sum_{j = a,b} \,\,
\frac{C_i\,A}{2}\, &e^{\frac{\sigma^2}{2\tau_i^2} - \frac{t}{\tau_i}} \left[ 1 + \text{erf}\left(\frac{t - \frac{\sigma^2}{\tau_i}}{\sigma \sqrt{2}} \right) \right] \, + \\
\frac{C_i\,(1-A)\,B_j}{2(\frac{\tau_j}{\tau_i} - 1)}\, &e^{\frac{\sigma^2}{2\tau_j^2} - 
\frac{t}{\tau_j}}\left[ 1 + \text{erf}\left(\frac{t - \frac{\sigma^2}{\tau_j}}{\sigma \sqrt{2}} \right) \right] \, - \\
\frac{C_i\,(1-A)\,B_j}{2(\frac{\tau_j}{\tau_i} - 1)}\, &e^{\frac{\sigma^2}{2\tau_i^2} - \frac{t}{\tau_i}}
\left[1 + \text{erf} \left( \frac{t - \frac{\sigma^2}{\tau_i}}{\sigma \sqrt{2}} \right) \right]\, .
\label{eq:analytic_model}
\end{split}
\end{equation}

\hspace{1ex}
\subsection{Scintillation Pulse Monte Carlo}\label{sec:WMC}
We developed a Monte Carlo (MC) that uses the analytical pulse shape model as an input and generates channel-level simulated signals. For a given scintillation pulse size, the number of photons that arrive at the top and bottom arrays are drawn randomly from a binomial distribution, using the top/bottom light collection asymmetry measured with the CH$_3$T calibration data. The number of photons in a single PMT channel is drawn from a binomial distribution, where we assume the probability of a photon landing in any given channel is $p = 1/61$. This is a good approximation for S1 light detection by the 61 PMTs in each array for events in the fiducial volume. The areas of each photon are independently drawn from the single-photon pulse area distribution, averaged across all PMTs. Each photon is then randomly assigned a detection time, drawn from the distribution in Eq. \ref{eq:analytic_model}. Photon template functions with the appropriate amplitudes and arrival times are added together to construct a simulated signal. Noise is added to this signal by adding sine waves with frequencies and amplitudes given by the measured noise power spectrum and random phases drawn uniformly on the interval $[0,2\pi]$. The bandwidth of the DAQ is therefore included in both the noise and the signal by using data-driven signal and noise distributions. The simulated waveform is then sampled at 100~MHz, with the starting point given by a uniform random number between 0 and 10~ns to simulate timing jitter due to the digitizer's sampling. This MC is used in the following sections to simulate scintillation pulses for error analyses and discrimination calculations.


\section{ER/NR data analysis and results}
\label{sec:analysis}

\subsection{Photon Time Spectrum}
\begin{figure}
    \includegraphics[width=0.5\textwidth]{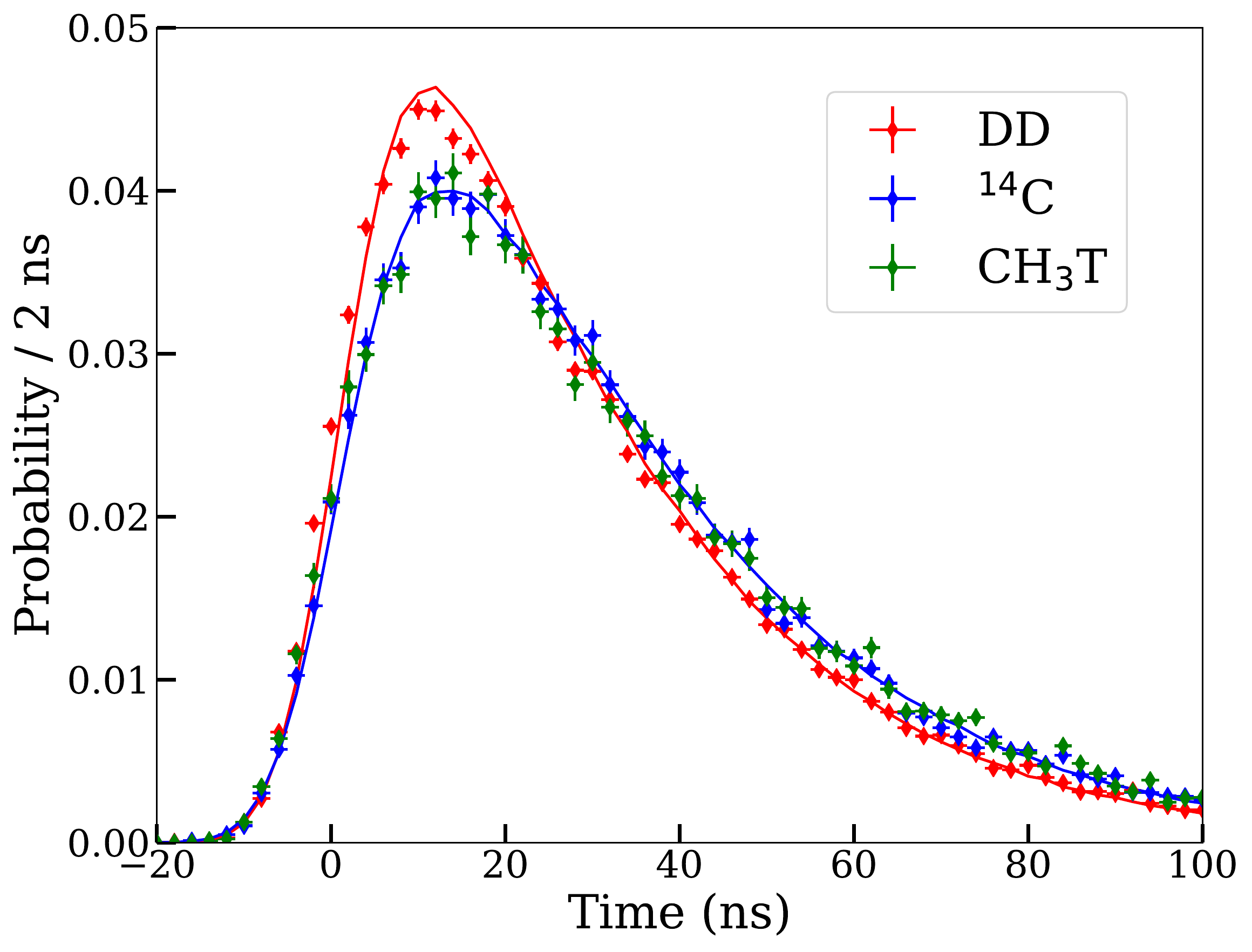}
    \caption{Average photon detection time spectra for events with pulse area between 40--50~phd ($\sim11$--13~keV$_{ee}$). The probability distribution of NR events, from DD neutron data, is shown by the red diamonds and the MC simulated NR time spectrum is shown in red. The measured distributions for ER events from tritium and $^{14}$C are shown in green and blue, respectively. The ER time spectrum generated from MC simulations is shown in blue. The time at which the pulse area reaches 5\% of the total, denoted T05 in the text, is used as $t=0$. The vertical uncertainties in each bin are calculated from Poisson statistics, while the horizontal error bars represent bin width.}
    \label{fig:example} 
\end{figure}

We study scintillation characteristics in ER/NR calibration data by constructing photon time spectra that are averaged over many events. Event selection is based on the dark matter search analyses: we study only single-scatter events, defined as events with a single S2 preceded by a single S1 within the maximum drift time (330~$\mu$s). To reduce the position dependence of optical transport, indicated by Fig.~\ref{fig:optical}, we select only events in a rectangular prism around the beam path of the neutron calibration source, at a median depth of $\sim$7.5~cm below the liquid surface. The average electric field in this region, calculated using COMSOL electrostatics simulation software~\cite{comsolRef}, is 410~V/cm. The times of detected photons in these events are corrected using the channel-to-channel time calibration and the direct-path travel time from the event site to the PMT is subtracted, and weighted by the $w_i$'s. To align time spectra from different events, we define a common reference of the sample at which the summed waveform crosses 5\% of the total pulse area (hereafter denoted T05). The photon time distribution for many events are used to produce average time spectra. As our analysis is focused on the true number of photons arriving in a given pulse, we use S1 pulse areas that are not corrected for position dependent effects in the LUX detector unless otherwise noted. 

Average time spectra for the three calibration sources with pulse area between 40--50~phd are shown by the data points in Fig.~\ref{fig:example}. The two electron recoil sources ($^{14}$C and tritium) show identical spectra, while the nuclear recoil source (DD neutrons) has a spectrum with a sharper peak in time. This difference is explored in the context of our analytical model in Section~\ref{subsec:fits}, and is used for ER/NR discrimination as discussed in more detail in Section~\ref{subsec:PSD}.  

The field- and position-dependence of the pulse shapes were studied in the tritium calibration data. The electric fields within the detector volume changed significantly between WS2013 and WS2014-16: in the former, the drift field was $\sim$180~V/cm throughout the detector, while in the latter, the drift field is highly non-uniform, varying from an average of $\sim$50~V/cm near the bottom to $\sim$400~V/cm near the top of the detector. For a fixed event position and fixed pulse area, we do not observe a significant difference in the average time spectra for any of the tritium calibrations from WS2013 or WS2014-16. We therefore conclude that there is no significant field-dependence within the limits of our sensitivity. We observe a depth-dependence in the average time spectra, consistent with the expectations from the optical transport model. Since photons from S1 pulses are preferentially detected in the bottom PMT array due to reflections at the liquid-gas interface, we attribute depth-dependence of the pulse shapes to the depth-dependence of the geometric efficiency of the bottom array. The effects of the depth-dependence on ER/NR discrimination is discussed in Section~\ref{subsec:PSD}.

\hspace{1ex}
\subsection{Fits to analytical model}
\label{subsec:fits}
The analytical model was fitted to the average time spectra in both ER and NR data to extract physical scintillation characteristics. To measure energy dependence in the model parameters, we fit the average time spectra binned by the reconstructed ER-equivalent energy (given in keV$_{ee}$),
\begin{equation}
E_{rec} = W\left(\frac{S1}{g_1} + \frac{S2}{g_2}\right)\, .
\label{eq:combined_energy}
\end{equation}
where $g_1$ and $g_2$ are the detector-specific gains for the $^{83m}$Kr-corrected $S1$ signals and $S2$ signals, and $W = 13.7\,\text{eV}$ is the average energy required to create either a scintillation photon or an ionization electron in liquid xenon \cite{LenardoNEST}. Over the period of the experiment, the parameters $g_1$ varied between 0.100~$\pm$~0.002 and 0.097~$\pm$~0.001~phd/photon, and $g_2$ varied between 18.92~$\pm$~0.82 and 19.72~$\pm$~2.39~phd/electron. For ER events, we separate the data into bins of 4~keV$_{ee}$ in $E_{rec}$ from 5--45~keV$_{ee}$. For NR events, we use bins of 2~keV$_{ee}$ from 5--17~keV$_{ee}$. To obtain the true energy of events contributing to each $E_{rec}$ bin, the distribution of recoil energies for each source is simulated using the NEST light and charge yield models tuned to LUX data, given in Ref.~\cite{LUXTritium} (for ER) and Ref.~\cite{LUXDD} (for NR). We report the mean and the $\pm1~\sigma$ of the simulated distributions as the true energy and its error. This analysis includes ER events with true energies from $\sim$5--46~keV and NR events with true energies from $\sim$25--74~keV. 

Several of the fit parameters are expected to remain constant across energies and particle types. The singlet time constant $\tau_1$ and the Gaussian fluctuation parameter $\sigma$ are expected to be the same across all energy bins and for both ER and NR data, as they are dependent solely on the scintillation physics of the Xe$^*_2$ dimer and the timing resolution, respectively. Similarly, the optical transport parameters ($A$, $B_a$, $\tau_a$, and $\tau_b$) depend solely on photon transport in LUX, and should be constant across all spectra for a fixed position inside the detector. The values used in these fits are given in Table~\ref{tab:transport} for the DD beam location, and are constant for both ER and NR spectra across all energies.

In contrast, the ratio of $C_1/C_3$ is allowed to vary independently for each energy bin in both ER and NR data. This allows our model to capture the difference in the singlet/triplet ratio between ER and NR events, as well as any possible dependence on recoil energy. We also allow $\tau_3$ to vary between ER and NR datasets to allow it to capture any small recombination effects, as discussed in Section~\ref{sec:modeling}. 

In addition to the timing effects built into the analytical model, the photon time spectra experience a spread due to statistical fluctuations in T05. These depend on the scintillation emission distribution and the total number of detected photons, and produce an additional smearing that could be mistaken for an energy dependence in the underlying time spectra. We model this effect using the MC to simulate events in each energy bin. The distribution of T05 in the appropriate energy bin is convolved with the analytical model before fitting to the measured average time spectra.

In order to fit all of these parameters with the appropriate constraints and correlations, we simultaneously fit the average time spectra at all energies using a global log-likelihood given by
\begin{equation}
\begin{split}
\text{log}(L) = &\left[\,\,\sum_{\text{ER,NR}}\,\, \sum_{i=1}^{N}\,\, 
\sum_{j=1}^{M} -\frac{(Y_j - Y_{calc})^2}{\sqrt{2}\,\,Y_j} \right]\, ,
\end{split}
\label{eq:likelihood}
\end{equation}
where $N$ is the number of bins in energy, $M$ is the number of bins in the average pulse timing distribution for each energy bin, $Y_j$ is the number of photons in bin $j$ of the timing distribution, and $Y_{calc}$ is the height of bin $j$, calculated by the model. The singlet and triplet times are constrained to vary within [0~ns,~10~ns] and [18~ns,~35~ns], respectively, to avoid degeneracy. We maximize log~$(L)$ using the Minuit optimizer class provided by the ROOT framework \cite{MinuitPaper}.

{\setlength{\tabcolsep}{1.4em}
\begin{table*}[t]
{\def\arraystretch{1.6}
 \caption{Summary of parameters used in fitting our photon time spectra. Expected values in column 2 come from the literature where appropriate: the predictions for the triplet and singlet times come from the average values in Ref.~\cite{MockNESTPulseShapes}, while the prediction for $C_1\tau_1/C_3\tau_3$ comes from the measurement in Ref.~\cite{XMASSPulseShape2016}. The expected value of $\sigma$ is calculated from Eq. \ref{eq:sigma}. The best-fit parameters are shown with $\pm 1~\sigma$ statistical uncertainties, calculated from the fit. The analysis systematic is computed by performing the same fit procedure on simulated data with known input parameters. The optical transport systematic uncertainty comes from varying the optical model used in the fit, i.e. fitting $A$, $B_a$, $\tau_a$, and $\tau_b$ to optical simulations with the $\pm1~\sigma$ extremes on optical parameters from Ref.~\cite{LUX_PRD}. }\label{tab:parameters}
\begin{tabular}{l c c c c c}
Parameter & Expected & Fit constraint & Best fit $\pm$ stat. & 
{\def\arraystretch{1.}\begin{tabular}[c]{@{}c@{}}Analysis\\sys.\end{tabular}} & 
{\def\arraystretch{1.}\begin{tabular}[c]{@{}c@{}}
Optical transport\\sys.\end{tabular}}\\  
\hline
\hline
$(C_1\tau_1)/(C_3\tau_3)$ & $\sim$0.1 (ER) & none & {0.042 $\pm$ 0.006 }&{$\pm$3.1\%  }&{$^{+75\%}_{-66\%}$  }\\ 
                      & none (NR)      & none & {0.269 $\pm$ 0.022   }&{$\pm$3.1\%  }&{$^{+20\%}_{-10\%}$  }\\
$\tau_1$   & 3.1 $\pm$ 0.7 ns         & 0-10~ns 
&{ 3.27 $\pm$ 0.66 ns }&{$\pm$1.0\% }&{$^{+11\%}_{-70\%}$ }\\ 
$\tau_3$   & 24 $\pm$ 1 ns (ER)     & 18-35~ns &{25.89 $\pm$ 0.06 ns }&{$\pm$1.9\% }&{$^{+0.5\%}_{-0.6\%}$ }\\
           & 24 $\pm$ 1 ns (NR)     & 18-35~ns &{23.97 $\pm$ 0.17 ns }&{$\pm$1.9\% }&{$^{+0.1\%}_{-1.1\%}$ }\\
$A$   & 0.0574        & fixed      & & & $^{+34\%}_{-9.1\%}$ \\
$B$   & 1.062        & fixed      & & & $^{+1.9\%}_{-1.0\%}$ \\
$\tau_a$ & 11.1 ns & fixed & & & $^{+17\%}_{-21\%}$ \\
$\tau_b$ & 2.70 ns & fixed & & & $^{+0\%}_{-8\%}$ \\
$\sigma$   & 3.2 ns         & none       &{ 3.84$\pm$ 0.09 ns }&{$\pm$1.1\% }&{$^{+1.2\%}_{-1.2\%}$ }\\
\end{tabular}}

\end{table*}}

The results of the global fit are summarized in Table \ref{tab:parameters}. The statistical errors are those returned by the fit routine. In addition, there are two main sources of systematic error: errors introduced in the analysis and fit procedure, and errors from uncertainties in the optical model used to produce $A$, $B_a$, $\tau_a$, and $\tau_b$. We quantify the first by performing the analysis on simulated events with known input parameters. Shifts between the input parameters and the reconstructed parameters are quoted as the systematic errors shown in the fifth column in Table \ref{tab:parameters}, and are $O$(2\%). The errors due to uncertainties in the optical model are quantified using the $\chi^2$ distribution of the optical model fit given in Ref.~\cite{LUX_PRD}. Of the seven free parameters in the optical model, we find that only two affect the photon transport times: the liquid xenon absorption length and the reflectivity of the teflon immersed in the liquid. We run new optical transport simulations allowing these parameters to vary along the $\Delta\chi^2$ = 8.18 (1$\sigma$) contour of the optical model parameter space. We then propagate these new simulated distributions into our pulse shape model and redo the analysis to extract new pulse shape parameters. We report the variations from the best-fit values as the systematic error, which is listed in the sixth column of Table~\ref{tab:parameters}. This is the dominant error in our analysis. 

\begin{figure}
{\includegraphics[width=0.45\textwidth]{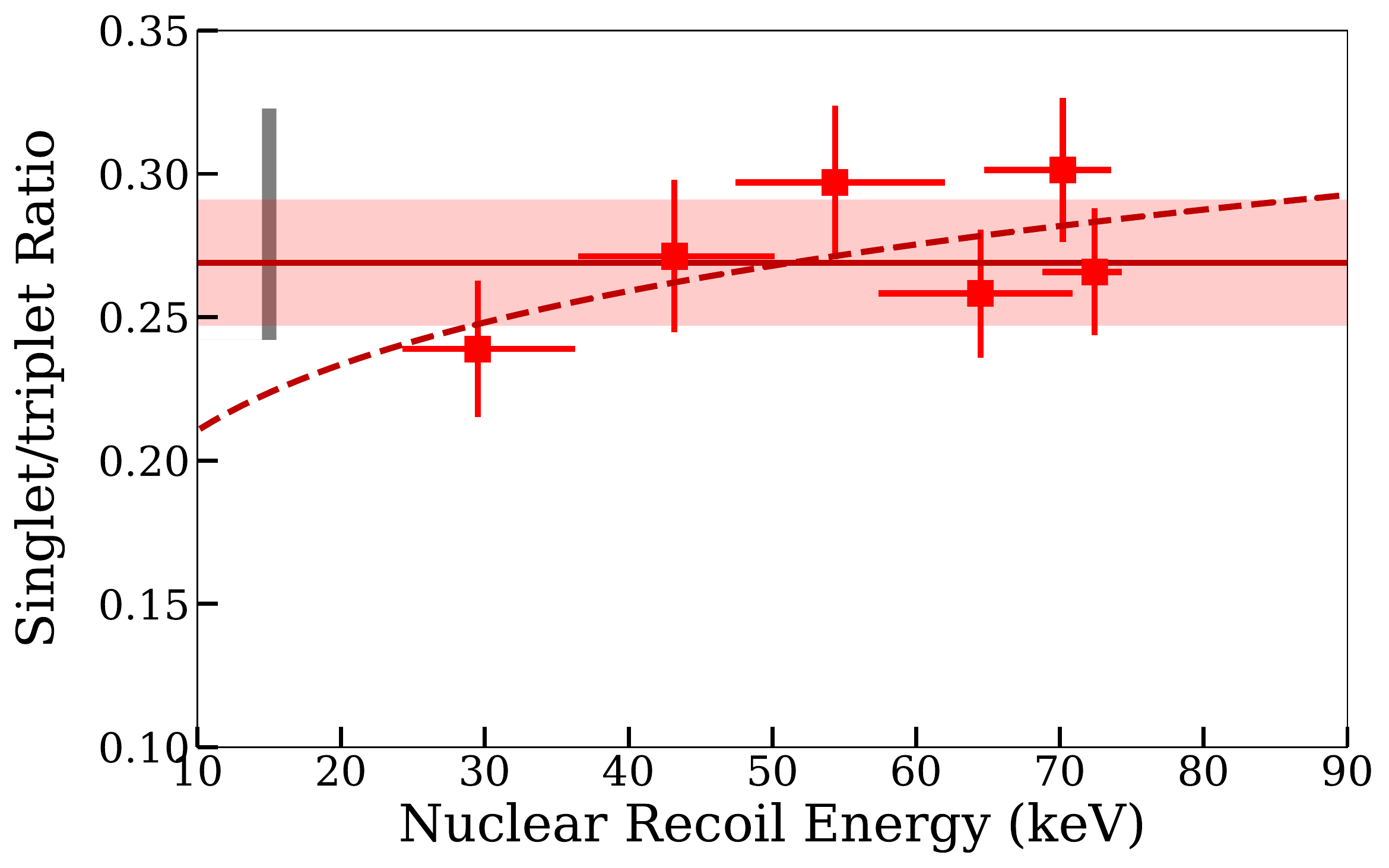}}
{\includegraphics[width=0.45\textwidth]{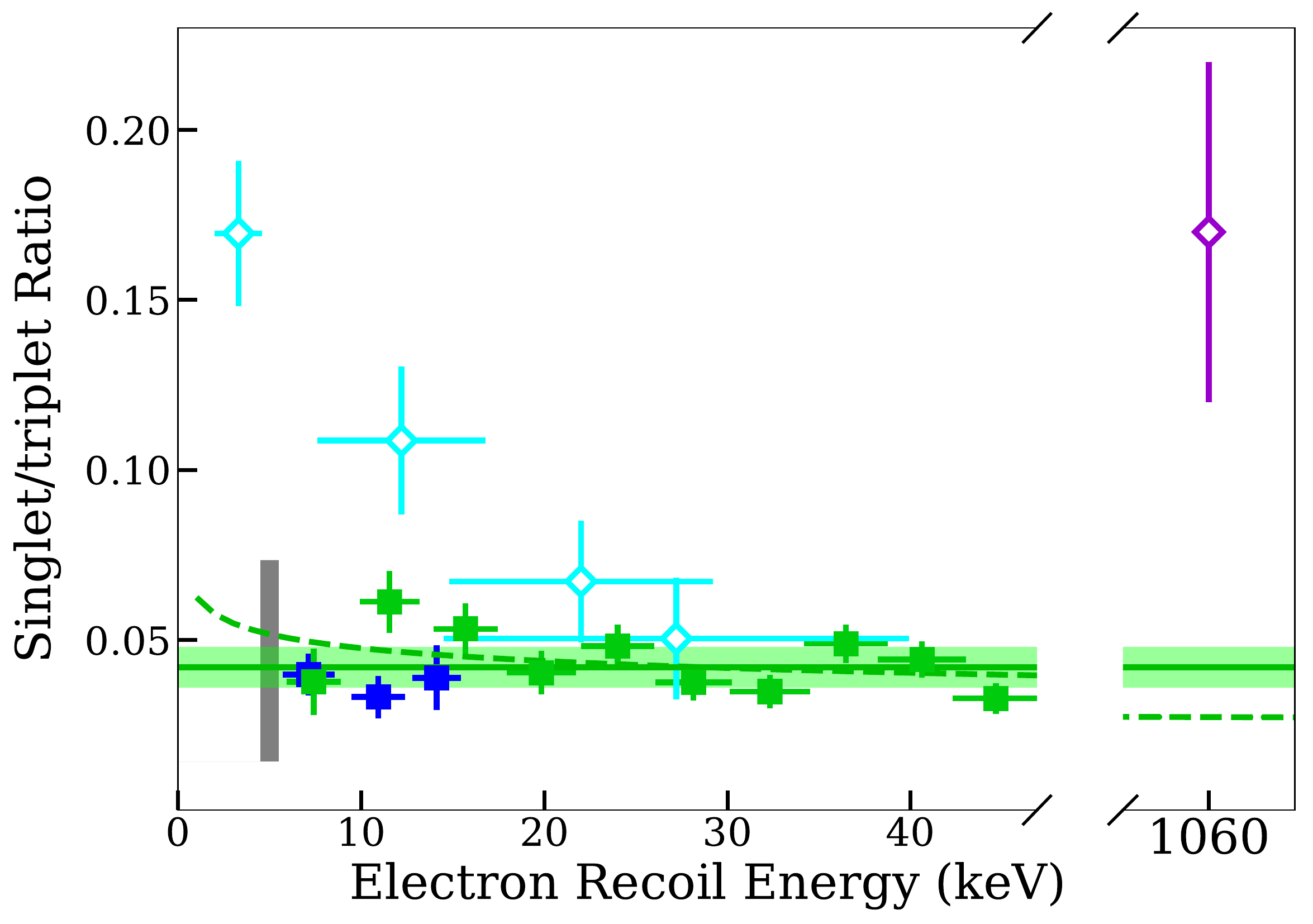}}
\caption{Singlet/triplet ratio ($C_1 \tau_1 / C_3 \tau_3$) measured for nuclear recoils (Top) and electron recoils (Bottom) using LUX calibration data. Only statistical uncertainties of the data are shown. Calibration sources are DD neutrons (red), tritium (blue), and $^{14}$C (green). Measurements in different energy bins are shown by the square points, and the best fit constant model by the solid line. The shaded region indicates the statistical uncertainty on the constant model. The shaded gray bar indicated the systematic uncertainty of the constant model. A power law is also fit to the data and is presented by the dashed line. We also show measurements of the ER singlet/triplet ratio at zero field from Ref.~\cite{XMASSPulseShape2016} (cyan diamonds), and a measurement using a $^{207}$Bi internal conversion source at 4~kV/cm from Ref.~\cite{KubotaRecombination} (purple diamond). In Ref.~\cite{XMASSPulseShape2016}, the singlet fraction (denoted $F_1$) is given rather than the singlet/triplet ratio. For direct comparison to this work we make the conversion $(C_1\tau_1)/(C_3\tau_3) = F_1 / (1 + F_1)$.}
\label{fig:singlet_triplet_ratios}
\end{figure}

The best-fit singlet/triplet ratios as a function of energy are shown in Fig.~\ref{fig:singlet_triplet_ratios}. For electron recoils we find $(C_1\tau_1)/(C_3\tau_3)$ = {$0.042 \pm 0.006 \,(\text{stat})\, ^{+0.092}_{-0.034}\, \text{(sys)}$}, averaged across all measured energies, which is lower than existing results in the literature. We note this is the first measurement of the singlet/triplet ratio with both a low energy ER source and an applied electric field. The energy dependence at zero field, measured by the XMASS collaboration (shown by cyan diamonds in Fig.~\ref{fig:singlet_triplet_ratios}) is correlated with a lengthening of the long time constant from 28~ns to 32~ns, which suggests that they are observing an increase in the recombination-related time constant with energy which are not explicitly accounted for in their model.  We do not observe an energy dependence in neither the time constant nor in the singlet/triplet ratio, consistent with the hypothesis that the applied electric field in our experiment suppresses recombination contributions to the pulse shape. For nuclear recoils, we find $(C_1\tau_1)/(C_3\tau_3)$ = {$0.269 \pm 0.022 \,(\text{stat})\, ^{+0.182}_{-0.083}\, \text{(sys)}$}, averaged across all energies probed in this analysis. The only analogous measurement in the literature uses recoiling fission fragments and finds $(C_1\tau_1)/(C_3\tau_3) = 1.6 \pm 0.2$, though in a vastly different energy regime at O(100~MeV) \cite{Hitachi1983PulseShape}.  Our result is therefore the first nuclear recoil singlet/triplet ratio measurement that is directly relevant for dark matter TPC experiments. 

We test for energy dependence of the singlet/triplet ratio by fitting both a constant value and a power law dependence, the latter given by $(C_1\tau_1)/(C_3\tau_3) = \alpha\,E^{\beta}$. Such an energy dependence is well-established in liquid argon~\cite{lippincott2008}, but has never been directly explored in xenon. For electron recoils, the best-fit values of the power law give $\alpha = 0.063~\textup{keV}^{-1}$ and $\beta = -0.12$. The $\chi^2$/d.o.f. for the constant and power law models are 16.6/9 (p = 0.06) and 13.7/8 (p = 0.09), respectively. For nuclear recoils, the best-fit values of the power law give $\alpha = 0.15~\textup{keV}^{-1}$ and $\beta = 0.15$. In this case, the $\chi^2$/d.o.f. for the constant and power law models are 4.6/5 (p = 0.47) and 3.2/4 (p = 0.52). We conclude that our data is statistically consistent with both models, and both are compared to data in Fig.~\ref{fig:singlet_triplet_ratios}, Fig.~\ref{fig:distributions}, and Fig.~\ref{fig:leakage} for completeness. 


Our best-fits of the triplet and singlet time constants, $\tau_1$ and $\tau_3$, agree with previously measured values. The expected values, listed in Table~\ref{tab:parameters}, are the error-weighted averages computed in Ref.~\cite{MockNESTPulseShapes} based on a survey of measurements in the literature. The only value in slight tension is the triplet time constant that we measure for electron recoils, which is higher than both the expected value and our best-fit for nuclear recoils. This is consistent with small recombination effects that are not accounted for in our model. If we assume $\tau_{3_{er}} = \tau_{3_{nr}} = 23.97$~ns and take the recombination time distribution derived in Ref.~\cite{KubotaRecombination} ($P(t) \propto [1 + (t/\tau_R)]^{-2}$), simulations reproduce our best-fit distribution with $\tau_R \approx 0.6$~ns. This expression for recombination time may not be directly applicable here, as it is derived by solving a diffusion equation with no applied electric field. However, we note that the qualitative agreement with the empirical prediction of $\tau_R = 0.7$~ns from Ref.~\cite{MockNESTPulseShapes} is encouraging. Regardless, our result for $\tau_{3_{er}}$ is still within the range of $\tau_3$ measurements available in the literature \cite{XMASSPulseShape2016,KubotaTriplet}, indicating that recombination plays a minor role in the pulse shapes for electron recoils in our experiment.

\hspace{1ex}
\section{Pulse Shape Discrimination}
\subsection{Prompt Fraction Discriminator}
\label{subsec:PSD}

\begin{figure}
    \centering
    \includegraphics[width=0.5\textwidth]{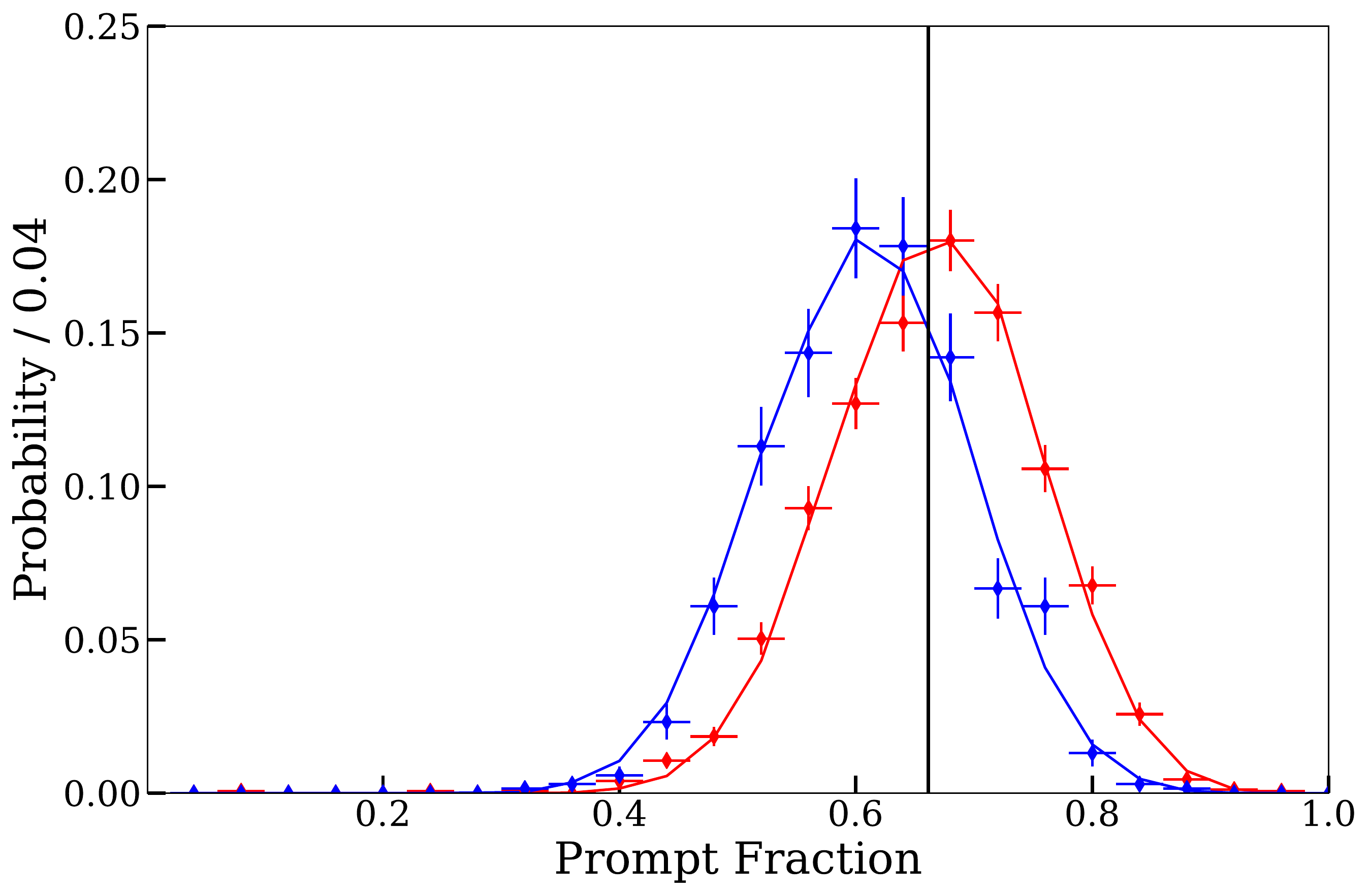}
    \caption{Normalized prompt fraction distributions for NR (red) and ER (blue) events with raw S1 pulse areas from 40--50~phd. Solid red and blue traces are the simulated NR and ER prompt fraction distributions for this pulse area. The vertical error bars indicate the Poisson uncertainty, while the horizontal error bars represent bin width. The solid black line indicates the median of the NR distribution, and the prompt fraction region above the DD median is considered to be the NR acceptance region. 26~$\pm$~2$\%$ of the ER distribution was found to lie within this region.} 
    \label{fig:dist} 
\end{figure}

To discriminate between ER and NR events we use a Prompt Fraction Discriminator (PFD), a standard technique which has been successfully adapted for use in other liquid xenon and liquid argon dark matter experiments \cite{XENON10InelasticDM, ZEPLIN_I, ZEPLIN3_IDM, DarksideFirstResult, XMASSPulseShape2016}. The parameter is defined as:
\begin{equation}
PF = \frac{\int_{t0}^{t1} \text{S1}(t) dt}{\int_{t2}^{t3} \text{S1}(t) dt} = \frac{\sum  \text{Prompt Photons}}{\sum \text{Total Photons}}\, .
\label{eq:psddef}
\end{equation}
The four variables $t0$, $t1$, $t2$, and $t3$, are allowed to vary independently in the range of $-$30 to 170~ns to minimize the leakage of ER events into the 50\% NR acceptance region (defined as everything above the NR median ($\overline{NR}$)). No additional constraints on these parameters were imposed and cases where $t0>t2$, etc. were explored. 

We apply an additional weighting scheme to avoid a bias in the optimization due to the energy dependence of the source. Since the yield at the calibration sources is energy dependent, we divide the data into 10 phd-wide bins. Each 10 phd bin is weighted equally when calculating the total leakage and is not weighted by the number of events in that particular bin. Doing so allows us to optimize the PFD for a ‘flat’ distribution in pulse area. 

To calculate the performance of the PFD, we separate the calibration datasets into two groups. Events in all datasets are randomly assigned to either a training or a testing group. Both groups contain 50\% of the data across all of the calibration campaigns and there is no statistically significant difference between their average detected photon time spectra, position or energy distributions. The training group of events are used to optimize our PFD. The optimized discriminant is then applied to the events in the testing group to quantify leakage and discrimination power below.

Carrying out the PFD optimization using the events in the $\it{training}$ group gives an optimal prompt window of $-$8 to 32~ns and total window of $-$14 to 134~ns. 
An example of the optimized PF values in the 40--50~phd bin, applied to events the $\it{testing}$ group, is presented in Fig.~\ref{fig:dist}. This PFD is also applied to events generated using the MC simulation and shows consistency with data. When this PFD is trained on the individual campaigns the optimal windows are found to vary up to 6~ns and are consistent with statistical fluctuations rather than real changes in the photon detection time spectrum.


\begin{figure*}
\centering
	\begin{subfigure}[b]{0.485\textwidth}
	\centering
	\includegraphics[width=\textwidth]{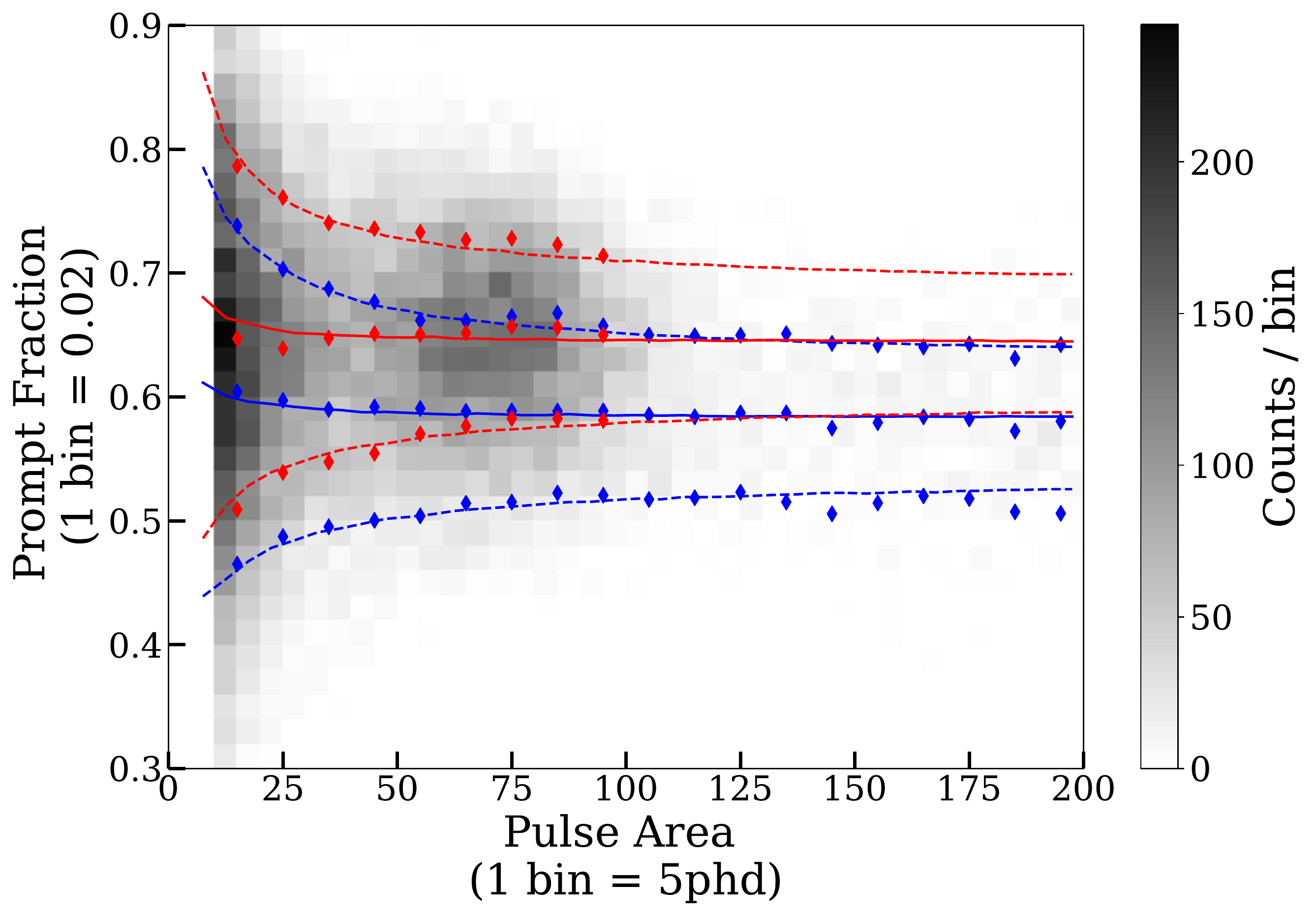}
	\caption[]
	{{\small NR data and the constant value model.}}    
	\end{subfigure}
\hfill
	\begin{subfigure}[b]{0.485\textwidth}  
	\centering 
	\includegraphics[width=\textwidth]{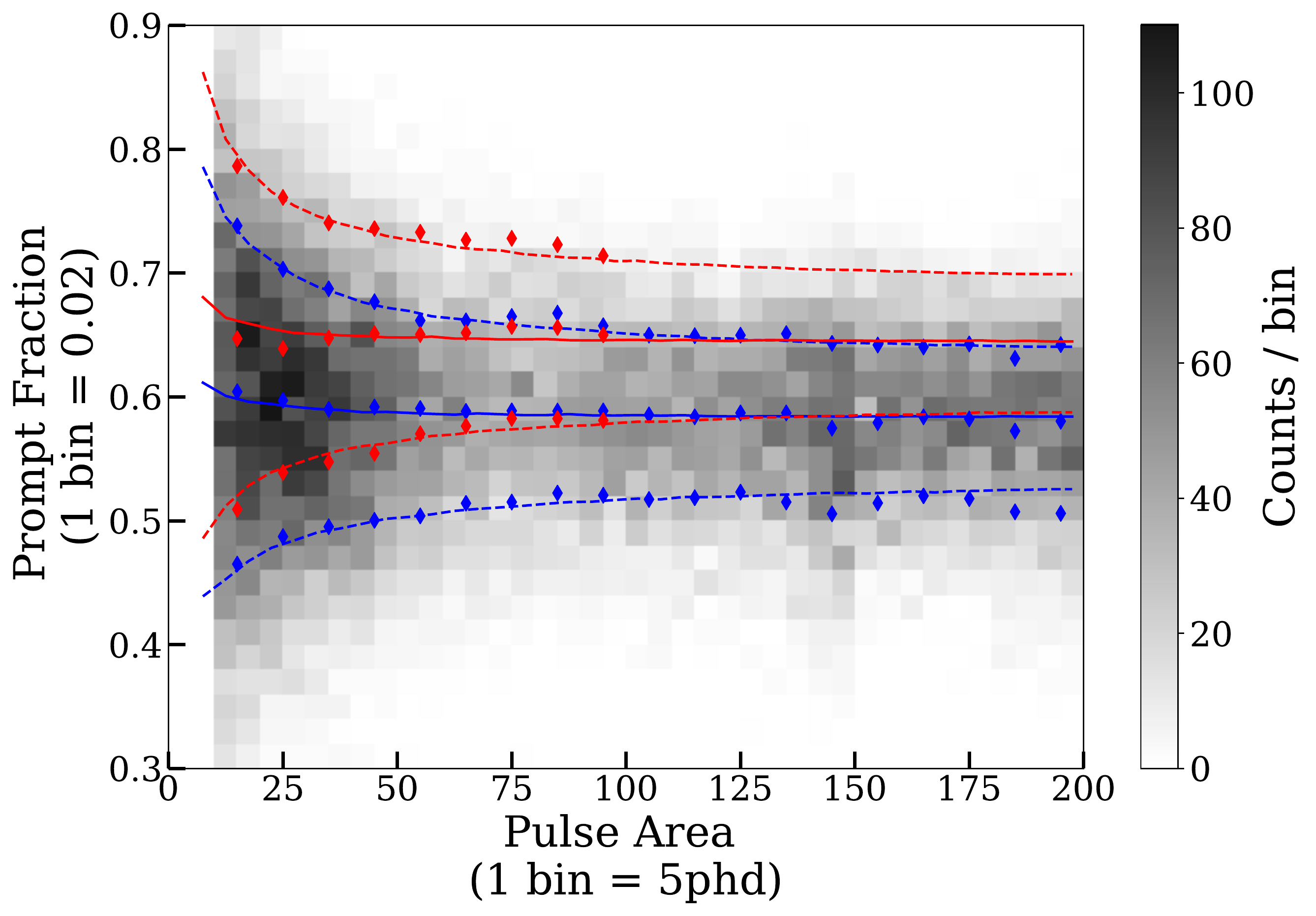}
	\caption[]
	{{\small ER data and the constant value model.}}    
    \end{subfigure}
\caption[]
{\small PF distribution for NR events \emph{(Left)}, and ER events \emph{(Right)}. The red and blue dots indicate the NR and ER bands respectively (median and median $\pm$34~percentile) calculated from data. The traces indicate the bands calculated from simulated data using the constant value model fitted to the singlet/triplet ratio (Fig.~\ref{fig:singlet_triplet_ratios}). The power law model produces bands similar to the constant value model in this energy region and is omitted for clarity. The solid traces indicate the medians and the dashed lines indicate the median $\pm$34~percentile. We define the region above the NR median as the NR acceptance region.} 
\label{fig:distributions}
\end{figure*}

\begin{figure}
    \centering
    \includegraphics[width=0.5\textwidth] {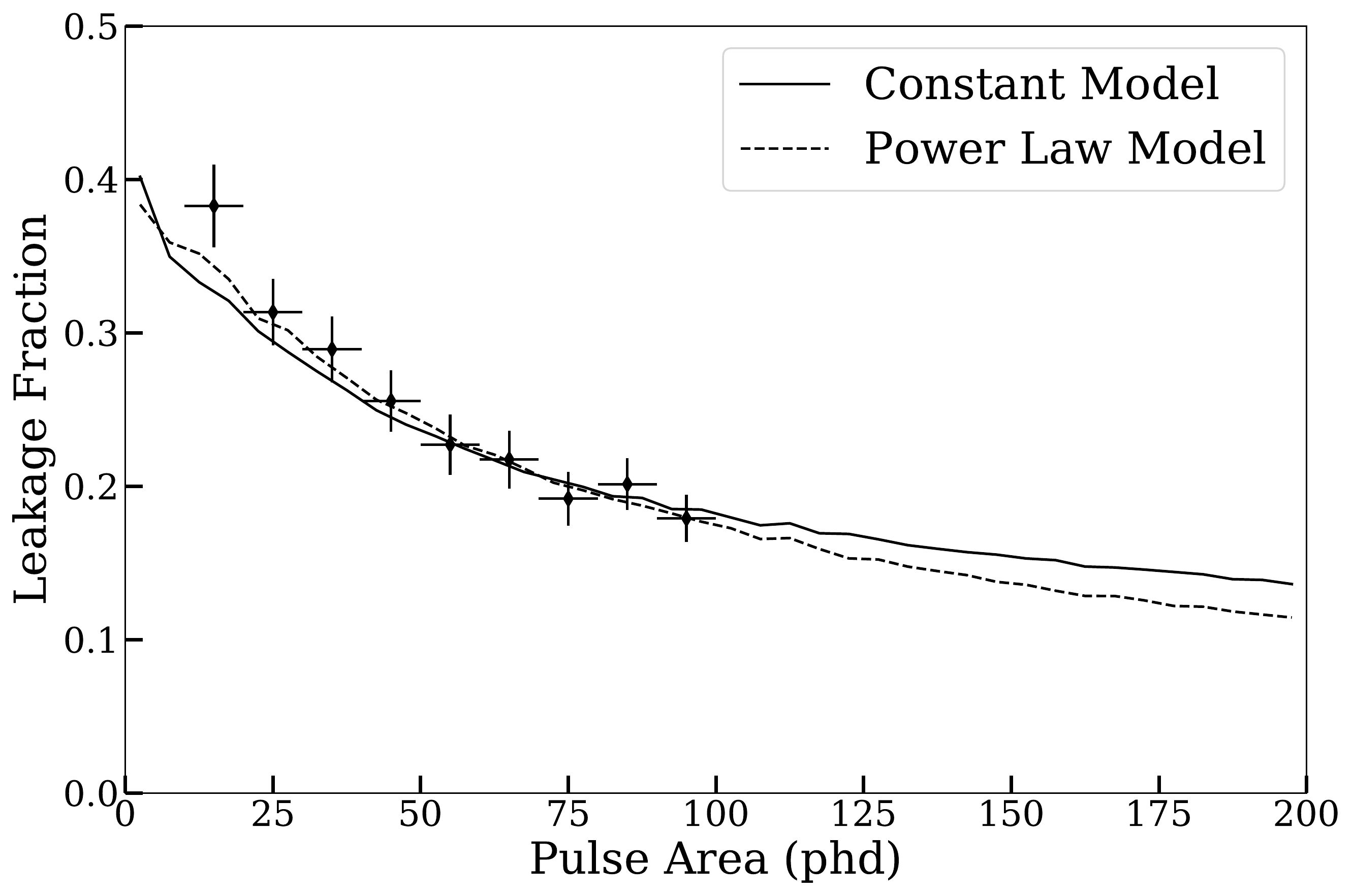}
    \caption{Fraction of ER events that leak into the NR region. The points show the leakage fraction obtained from the data while the solid trace indicates the leakage fraction calculated from simulation. The NR median is used to define the $\sim$50\% NR acceptance region. The horizontal error bars indicate the bin widths. The counting errors from Poisson statistics and the errors in NR median are added in quadrature to calculate the error in leakage fraction. For both data and simulations, the same PFD windows are used.} 
    \label{fig:leakage}
\end{figure}

Figure~\ref{fig:distributions} shows the PF vs. pulse area distributions calculated from DD, CH$_3$T and $^{14}$C calibration data. At small pulse areas, there is a large spread in PF due to the low photon statistics in each event. At larger pulse areas, more photons are reconstructed and the photon time spectra for individual events will appear more ER or NR-like, reducing the spread in PF. This effect provides improved ER/NR discrimination at higher energies. 
We also use the MC code to simulate prompt fraction values for both ER and NR, using the best-fit parameters found in Section~\ref{subsec:fits} to model the underlying scintillation and propagation physics. The bands in Figure~\ref{fig:distributions} are produced assuming a constant singlet/triplet ratio. We see that the simulated distributions match the data well.

Fig.~\ref{fig:leakage} shows the fraction of ER events that leak into the 50\% NR acceptance region. The DD calibrations were used to compute the 50\% NR acceptance region while the $^{14}$C and CH$_3$T calibrations were used to calculate the leakage into this region. For the distributions shown, the leakage in the lowest energy bin of 10--20~phd is 39.4~$\pm$~2.7\%. In the 40--50~phd bin, the highest bin used in the WS2013 and WS2014-16 analysis, the leakage fraction is reduced to 26.1~$\pm$~2.0\%. The leakage fraction continues to decrease at higher energies.

We also use calibration data at different depths to study the vertical position dependence on our discrimination power. At greater depths in the detector, the PFs move to larger values as more photons are detected at the bottom PMTs with less scattering. This geometric affect applies to both the ER and NR events and causes both bands to move by a similar value in PF for a given number of detected photons. As a result, we do not measure any significant depth dependence in the leakage fraction. While some depth-dependence is expected due to less scatter in the scintillation distributions, simulations indicate that this effect only changes the leakage fraction 0.6\% throughout the LUX fiducial volume, well within our statistical uncertainty. 
The overall simulated leakage fraction for a flat distribution up to 200~phd is 22.9\%.

\hspace{1ex}
\subsection{Two Parameter Discrimination for Dark Matter Searches}

{\setlength{\tabcolsep}{1.4em}
\begin{table*}[t]
{\def\arraystretch{1.6}
 \caption{Summary of ER leakage into the NR acceptance region using different methods. The data presented here are from all the DD, CH$_3$T and $^{14}$C calibrations from the volume around the DD beam. The average and $\pm1\,\sigma$ nuclear recoil energy for each bin is simulated for a flat energy spectrum using a NEST model tuned to LUX NR calibration data \cite{LUXDD}. In all cases the leakage is defined as the number of ER events that occur within the NR acceptance region. The errors indicated are from Poisson statistics. Column 3 presents the charge-to-light yield discriminator from Ref.~\cite{LUXRun4PRL} applied to this calibration data. The charge-to-light discriminator above 50~phd is subject of study. Column 6 presents performance of the PFD calculated from simulations using a linear singlet-to-triplet ratio.}\label{tab:disc}
\begin{tabular}{ c c c c c c c}
{\def\arraystretch{1.}\begin{tabular}[c]{@{}c@{}}S1 Pulse\\Area [phd]\end{tabular}}&
Energy [keV$_{nr}$] &
Log$_{10}$(S2/S1) [\%] &
{\def\arraystretch{1.}\begin{tabular}[c]{@{}c@{}}PFD\\Data [\%]\end{tabular}} &
{\def\arraystretch{1.}\begin{tabular}[c]{@{}c@{}}PFD\\Simulation [\%]\end{tabular}} &
{\def\arraystretch{1.}\begin{tabular}[c]{@{}c@{}}Two\\Parameter Data [\%]\end{tabular}}\\
\hline
\hline
10-20 & $16.1^{+4.8}_{-4.1}$ & 0.5 $\pm$ 0.2 & 39.3 $\pm$ 2.7 & 32.7 & 0.4 $\pm$ 0.2 \\
20-30 & $23.7^{+5.3}_{-4.4}$ &  0.4 $\pm$ 0.2 & 31.3 $\pm$ 2.2 & 29.4 & 0.3 $\pm$ 0.1 \\
30-40 & $31.1^{+6.0}_{-5.1}$ & 0.4 $\pm$ 0.2 & 28.9 $\pm$ 2.2 & 26.9 & 0.2 $\pm$ 0.1 \\
40-50 & $38.4^{+6.2}_{-5.8}$ & 0.3 $\pm$ 0.2 & 25.6 $\pm$ 2.0 & 24.5 & 0.1 $\pm$ 0.1 \\
50-60 & $45.0^{+7.1}_{-6.2}$ & & 22.7 $\pm$ 2.0 & 22.9 & 0.1 $\pm$ 0.1 \\
60-70 & $51.8^{+7.6}_{-6.6}$ & & 21.7 $\pm$ 1.9 & 21.3 & 0.0 $\pm$ 0.1 \\
70-80 & $58.9^{+7.7}_{-7.4}$ & & 19.2 $\pm$ 1.8 & 20.2 & 0.0 $\pm$ 0.1 \\
80-90 & $65.4^{+8.5}_{-7.5}$ & & 20.1 $\pm$ 1.7 & 19.3 & 0.1 $\pm$ 0.1 \\
90-100 & $72.2^{+8.7}_{-8.0}$ & & 17.9 $\pm$ 1.5 & 18.5 & 0.0 $\pm$ 0.1 \\
\\
\end{tabular}}
\end{table*}}

\begin{figure}
    \includegraphics[width=0.5\textwidth]{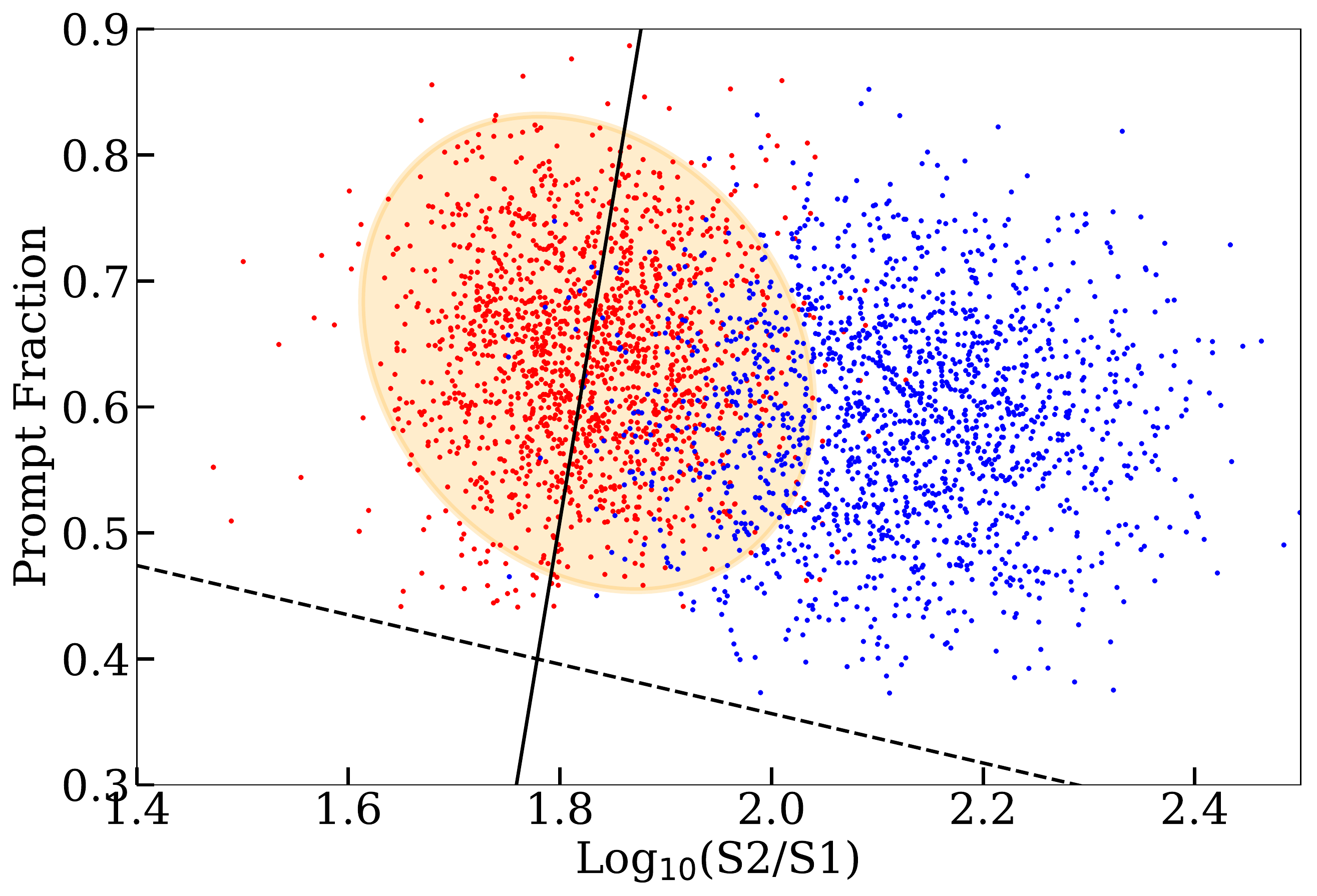}
    \includegraphics[width=0.5\textwidth]{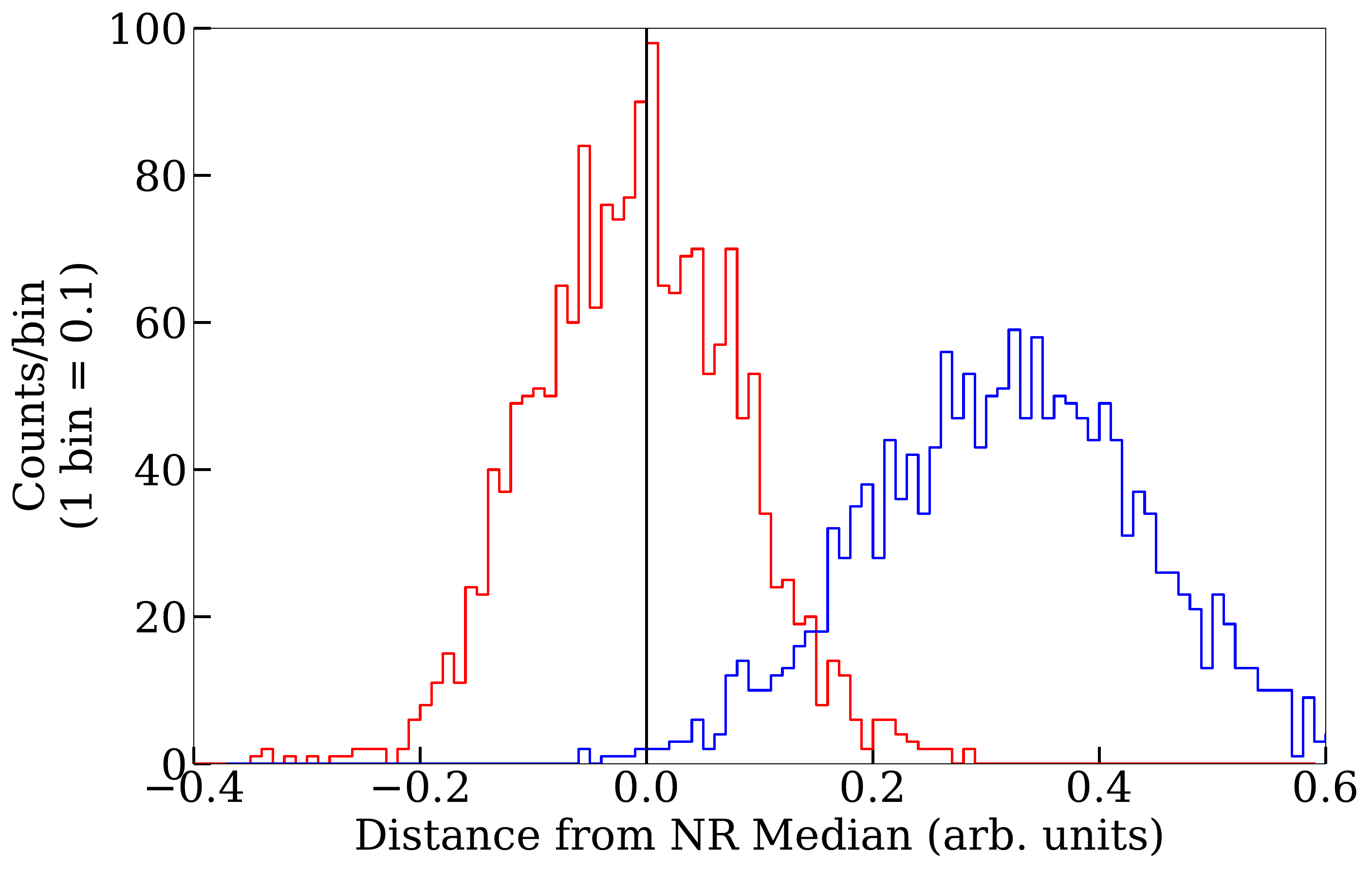}
    \caption{\emph{(Top)} Two parameter event discrimination using PFD and log$_{10}$($S2/S1$). The distribution contains NR (red) and ER (blue) events with S1 pulse areas between 40--50~phd. The yellow ellipse indicates the 90\% NR region and the solid black line indicates the discriminator. These events are projected on to the dashed black axis perpendicular to the NR Median. \emph{(Bottom)} The distance of these events from the NR median is presented. The region to the left of the solid black line is defined as the NR acceptance region and is chosen to preserve 50\% of the ellipse's NR region while minimizing the number of ER events in this region.} 
    \label{fig:twopara}
\end{figure}

The PFD can be used in conjunction with the charge-to-light ratio (log$_{10}$(S2/S1)) to develop a two-dimensional discriminant against ER backgrounds in LUX. This is shown in Fig.~\ref{fig:twopara} for events with pulse areas between 40--50~phd. An elliptical region, centered at the median values of the NR distribution, is chosen to include $\sim$90\% of the total distribution. In this two dimensional space, a line passing through the median of the NR population defines a linear cut to discriminate between the two populations of events. The region above this linear cut, away from the ER population, is defined as the NR acceptance region. In this manner, $\sim$50~\% of the NR acceptance region is preserved by the ellipse and the vertical cut.

For each 10~phd bin in pulse area, the three free parameters of the ellipse (inclination and two radii) and the one free parameter of the linear cut ($x$ or $y$-intercept) vary to minimize the ER leakage into the NR acceptance region. To test leakage, a line is drawn through the center of the ellipse with either a $x$-intercept between 1.25 and 2.75 or a $y$-intercept between 0 and 1. In this way, a cut that closely resembles just the charge-to-light discrimination is also tested. When this method is applied to 10~phd wide bins between 10--100~phd, the ellipse's inclination and $x$-intercept with the least leakage are often very similar. For each of these bins, the optimal inclination of the NR population and $x$-intercept and are close to 30$^{\circ}$ counter-clockwise w.r.t. the vertical. and 1.7, respectively. As the pulse area of the event gets smaller, the distribution of possible PF and charge-to-light increase, and thus the optimal radii for the ellipses vary to capture this change. 

For the example shown in Fig.~\ref{fig:twopara} for 40--50~phd, the measured number of ER events appearing in the NR acceptance regions is reduced compared to just the log$_{10}$($S2/S1$). Using the log$_{10}$($S2/S1$) bands for this population of events~\cite{LUXRun4PRL}, an overall ER leakage of 0.4~$\pm$~0.1~\% is achieved. When the log$_{10}$($S2/S1$) and PFD are combined to produce a new discrimination parameter, as described above, the overall ER leakage reduces to 0.3~$\pm$~0.1~\%. The comparison of the various discrimination methods are presented in Table~\ref{tab:disc}.

\hspace{1ex}
\section{Conclusion}\label{sec:conc}

We have described an analysis of liquid xenon scintillation pulse shapes and the discrimination power in the LUX dark matter experiment. We have developed software that allows for the precise reconstruction of photon detection times within a pulse with an accuracy of 3.8~ns. LUX calibration data from DD, CH$_3$T, and $^{14}$C sources are used to characterize the photon detection time spectra for NR and ER events at various depths in LUX. Average time spectra are fitted with an analytical model to extract singlet-to-triplet ratios and singlet and triplet decay times. It is found that the singlet-to-triplet ratio for ER events is consistent with the literature within errors. We have made a first measurement of the NR singlet-to-triplet ratio at low energies and a non-zero applied electric field. Different $\tau_3$ time constants are found for ER and NR events. We interpret this as residual recombination timing effects, which are not included in our model, adding a small smearing to the ER pulse shape which gets captured by the $\tau_3$ parameter in our fits. These measurements and the reconstructed physical properties of xenon scintillation are relevant for liquid xenon dark matter search experiments, and can inform simulation packages, such as NEST, that are used by the community to compute event distributions in current and future experiments.

The template-fitting timing algorithm is applied to calibration data to construct photon detection time spectra. The difference between ER and NR time spectra is exploited to formulate a ratio of prompt to total photons to discriminate ER and NR events. This discrimination parameter (PF) is optimized, using a training data set, to minimize the leakage of ER events into the $\sim$50\% NR acceptance region. The photon detection time and the prompt fraction distributions are shown to agree with those generated from the MC simulations using the best-fit analytical model, allowing us to extrapolate to energy regions where no calibration data is available. The discrimination power of the PFD improves with energy. For a flat distribution of events in the WS2013 and WS2014-16 analysis region from 10--50~phd, the ER leakage is found to be 31.3~\%. Between 10--200~phd, the average leakage is 25.2~\% . 

In the two-dimensional parameter space composed of charge-to-light ratio (log$_{10}$($S2/S1$)) and PF, an improved discriminator is developed. This discriminator is required to preserve $\sim$50~\% NR acceptance while reducing the ER leakage into the region. Over the WS2013 and WS2014-16 analysis region of 10--50~phd, the ER leakage, measured using the charge-to-light discriminator, is 0.4~$\pm$~0.1~\% and reduces to 0.3~$\pm$~0.1~\%, measured using the two parameter discriminator. Given the increase in photon statistics in both the singlet and triplet scintillation, we expect the discrimination power would increase at higher energies. If this expectation holds true after conducting appropriate calibrations, this approach would be an attractive background reduction technique for dark matter searches looking for nuclear recoils at energies higher than the traditional WIMP search. Examples of these dark matter searches include models in which dark matter scatters inelastically, or with a momentum-dependent cross-section. Using the parameters from the analytical model, the pulse shape discrimination bands can be extrapolated out to higher energies than accessible by calibration sources, or can be extrapolated to assess the pulse shape discrimination power in future liquid xenon experiments.


\begin{acknowledgments}
This work was partially supported by the U.S. Department of Energy (DOE) under award numbers DE-AC02-05CH11231, DE-AC05-06OR23100, DE-AC52-07NA27344, DE-FG01-91ER40618, DE-FG02-08ER41549, DE-FG02-11ER41738, DE-FG02-91ER40674, DE-FG02-91ER40688, DE-FG02-95ER40917, DE-NA0000979, DE-SC0006605, DE-SC0010010, and DE-SC0015535; the U.S. National Science Foundation under award numbers PHY-0750671, PHY-0801536, PHY-1003660, PHY-1004661, PHY-1102470, PHY-1312561, PHY-1347449, PHY-1505868, and PHY-1636738; the Research Corporation grant RA0350; the Center for Ultra-low Background Experiments in the Dakotas (CUBED); and the South Dakota School of Mines and Technology (SDSMT). LIP-Coimbra acknowledges funding from Funda\c{c}\~{a}o para a Ci\^{e}ncia e a Tecnologia (FCT) through the project-grant PTDC/FIS-NUC/1525/2014. Imperial College and Brown University thank the UK Royal Society for travel funds under the International Exchange Scheme (IE120804). The UK groups acknowledge institutional support from Imperial College London, University College London and Edinburgh University, and from the Science \& Technology Facilities Council for PhD studentships ST/K502042/1 (AB), ST/K502406/1 (SS) and ST/M503538/1 (KY). The University of Edinburgh is a charitable body, registered in Scotland, with registration number SC005336.

This research was conducted using computational resources and services at the Center for Computation and Visualization, Brown University, and also the Yale Science Research Software Core. The $^{83}$Rb used in this research to produce $^{83\mathrm{m}}$Kr was supplied by the United States Department of Energy Office of Science by the Isotope Program in the Office of Nuclear Physics.

We gratefully acknowledge the logistical and technical support and the access to laboratory infrastructure provided to us by SURF and its personnel at Lead, South Dakota. SURF was developed by the South Dakota Science and Technology Authority, with an important philanthropic donation from T. Denny Sanford, and is operated by Lawrence Berkeley National Laboratory for the Department of Energy, Office of High Energy Physics.

\end{acknowledgments}


\begin{thebibliography}{0}%
\makeatletter
\providecommand \@ifxundefined [1]{%
 \@ifx{#1\undefined}
}%
\providecommand \@ifnum [1]{%
 \ifnum #1\expandafter \@firstoftwo
 \else \expandafter \@secondoftwo
 \fi
}%
\providecommand \@ifx [1]{%
 \ifx #1\expandafter \@firstoftwo
 \else \expandafter \@secondoftwo
 \fi
}%
\providecommand \natexlab [1]{#1}%
\providecommand \enquote  [1]{``#1''}%
\providecommand \bibnamefont  [1]{#1}%
\providecommand \bibfnamefont [1]{#1}%
\providecommand \citenamefont [1]{#1}%
\providecommand \href@noop [0]{\@secondoftwo}%
\providecommand \href [0]{\begingroup \@sanitize@url \@href}%
\providecommand \@href[1]{\@@startlink{#1}\@@href}%
\providecommand \@@href[1]{\endgroup#1\@@endlink}%
\providecommand \@sanitize@url [0]{\catcode `\\12\catcode `\$12\catcode
  `\&12\catcode `\#12\catcode `\^12\catcode `\_12\catcode `\%12\relax}%
\providecommand \@@startlink[1]{}%
\providecommand \@@endlink[0]{}%
\providecommand \url  [0]{\begingroup\@sanitize@url \@url }%
\providecommand \@url [1]{\endgroup\@href {#1}{\urlprefix }}%
\providecommand \urlprefix  [0]{URL }%
\providecommand \Eprint [0]{\href }%
\providecommand \doibase [0]{http://dx.doi.org/}%
\providecommand \selectlanguage [0]{\@gobble}%
\providecommand \bibinfo  [0]{\@secondoftwo}%
\providecommand \bibfield  [0]{\@secondoftwo}%
\providecommand \translation [1]{[#1]}%
\providecommand \BibitemOpen [0]{}%
\providecommand \bibitemStop [0]{}%
\providecommand \bibitemNoStop [0]{.\EOS\space}%
\providecommand \EOS [0]{\spacefactor3000\relax}%
\providecommand \BibitemShut  [1]{\csname bibitem#1\endcsname}%
\let\auto@bib@innerbib\@empty
\end{thebibliography}%


\begin{thebibliography}{99}

\bibitem{LUXRun4PRL}
D.~S. Akerib, S.~Alsum, H.~M. Ara\'ujo, {\em et~al.}, ``{Results from a Search
  for Dark Matter in the Complete LUX Exposure},'' {\em Phys. Rev. Lett.},
  vol.~118, p.~021303, Jan 2017.

\bibitem{PandaXSpinIndependent2016}
A.~Tan, M.~Xiao, X.~Cui, {\em et~al.}, ``{Dark Matter Results from First 98.7
  Days of Data from the PandaX-II Experiment},'' {\em Phys. Rev. Lett.},
  vol.~117, p.~121303, Sep 2016.

\bibitem{XENON100Result}
E.~Aprile, M.~Alfonsi, K.~Arisaka, {\em et~al.}, ``{Dark Matter Results from
  225 Live Days of {XENON}100 Data},'' {\em Phys. Rev. Lett.}, vol.~109,
  p.~181301, Nov 2012.

\bibitem{MartinExcimers}
M.~Martin, ``{Exciton self-trapping in rare-gas crystals},'' {\em J. Chem.
  Phys.}, vol.~54, no.~8, pp.~3289--3299, 1971.

\bibitem{FUJII2015293}
K.~Fujii, Y.~Endo, Y.~Torigoe, {\em et~al.}, ``{High-accuracy measurement of
  the emission spectrum of liquid xenon in the vacuum ultraviolet region},''
  {\em Nucl. Instr. Meth. Phys. Res. A}, vol.~795, pp.~293 -- 297, 2015.

\bibitem{KubotaTriplet}
S.~Kubota, M.~Hishida, and J.~Raun, ``{Evidence for a triplet state of the
  self-trapped exciton states in liquid argon, krypton and xenon},'' {\em J.
  Phys. C}, vol.~11, no.~12, p.~2645, 1978.

\bibitem{Hitachi1983PulseShape}
A.~Hitachi, T.~Takahashi, N.~Funayama, {\em et~al.}, ``Effect of ionization
  density on the time dependence of luminescence from liquid argon and xenon,''
  {\em Phys. Rev. B}, vol.~27, pp.~5279--5285, May 1983.

\bibitem{KubotaRecombination}
S.~Kubota, M.~Hishida, M.~Suzuki, and J.~Ruan(Gen), ``Dynamical behavior of
  free electrons in the recombination process in liquid argon, krypton, and
  xenon,'' {\em Phys. Rev. B}, vol.~20, pp.~3486--3496, Oct 1979.

\bibitem{Dawson2005}
J.~Dawson, A.~Howard, D.~Akimov, {\em et~al.}, ``A study of the scintillation
  induced by alpha particles and gamma rays in liquid xenon in an electric
  field,'' {\em Nucl. Instr. Meth. Phys. Res. A}, vol.~545, no.~3, pp.~690 --
  698, 2005.

\bibitem{Akimov2002}
D.~Akimov, A.~Bewick, D.~Davidge, {\em et~al.}, ``Measurements of scintillation
  efficiency and pulse shape for low energy recoils in liquid xenon,'' {\em
  Phys. Lett. B}, vol.~524, no.~3--4, pp.~245 -- 251, 2002.

\bibitem{XMASSPulseShape2016}
H.~Takiya, K.~Abe, K.~Hiraide, {\em et~al.}, ``{{A measurement of the time
  profile of scintillation induced by low energy gamma-rays in liquid xenon
  with the XMASS-I detector}},'' {\em Nucl. Instr. Meth. Phys. Res. A},
  vol.~834, pp.~192 -- 196, 2016.

\bibitem{MockNESTPulseShapes}
J.~Mock, N.~Barry, K.~Kazkaz, {\em et~al.}, ``{Modeling Pulse Characteristics
  in Xenon with NEST},'' {\em J. Instrum.}, vol.~9, p.~T04002, 2014.

\bibitem{Kwong2010}
J.~Kwong, P.~Brusov, T.~Shutt, {\em et~al.}, ``Scintillation pulse shape
  discrimination in a two-phase xenon time projection chamber,'' {\em Nucl.
  Instr. Meth. Phys. Res. A}, vol.~612, no.~2, pp.~328 -- 333, 2010.

\bibitem{Ueshima2011}
K.~{Ueshima}, K.~{Abe}, K.~{Hiraide}, {\em et~al.}, ``{Scintillation-only based
  pulse shape discrimination for nuclear and electron recoils in liquid
  xenon},'' {\em Nucl. Instr. Meth. Phys. Res. A}, vol.~659, pp.~161--168, Dec.
  2011.

\bibitem{DAMAXe}
R.~Bernabei, P.~Belli, F.~Montecchia, {\em et~al.}, ``New limits on particle
  dark matter search with a liquid xenon target-scintillator,'' {\em Phys.
  Lett. B}, vol.~436, no.~3, pp.~379 -- 388, 1998.

\bibitem{ZEPLIN_I}
G.~Alner, H.~Ara{\'u}jo, G.~Arnison, {\em et~al.}, ``{First limits on nuclear
  recoil events from the ZEPLIN I galactic dark matter detector},'' {\em
  Astropart. Phys.}, vol.~23, pp.~444--462, June 2005.

\bibitem{XENON10InelasticDM}
J.~Angle, E.~Aprile, F.~Arneodo, {\em et~al.}, ``Constraints on inelastic dark
  matter from {XENON}10,'' {\em Phys. Rev. D}, vol.~80, p.~115005, Dec 2009.

\bibitem{XMASSLightWIMPSearch}
K.~Abe, K.~Hieda, K.~Hiraide, {\em et~al.}, ``{Light WIMP search in XMASS},''
  {\em Phys. Lett. B}, vol.~719, no.~1–3, pp.~78 -- 82, 2013.

\bibitem{XENON1T_PhysicsReach}
E.~Aprile, J.~Aalbers, F.~Agostini, {\em et~al.}, ``{Physics reach of the
  {XENON1T} dark matter experiment},'' {\em J. Cosmol. Astropart. Phys.},
  vol.~2016, no.~04, p.~027, 2016.

\bibitem{LZ_CDR}
D.~A. et~al., ``{The LUX-ZEPLIN Conceptual Design Report (CDR)},'' {\em
  arXiv:1509.02910}, 2015.

\bibitem{Akerib20121}
D.~Akerib, X.~Bai, S.~Bedikian, {\em et~al.}, ``{Data acquisition and readout
  system for the LUX dark matter experiment},'' {\em Nucl. Instr. Meth. Phys.
  Res. A}, vol.~668, pp.~1 -- 8, 2012.

\bibitem{NESTpaper}
M.~Szydagis, N.~Barry, K.~Kazkaz, {\em et~al.}, ``{NEST: A Comprehensive Model
  for Scintillation Yield in Liquid Xenon},'' {\em J. Instrum.}, vol.~6,
  p.~P10002, 2011.

\bibitem{Fitzpatrick2012_EFT}
A.~L. Fitzpatrick, W.~Haxton, E.~Katz, N.~Lubbers, and Y.~Xu, ``{The Effective
  Field Theory of Dark Matter Direct Detection},'' {\em JCAP}, vol.~1302,
  p.~004, 2013.

\bibitem{Bramante_Inelastic}
J.~Bramante, P.~J. Fox, G.~D. Kribs, and A.~Martin, ``The inelastic frontier:
  Discovering dark matter at high recoil energy,'' {\em arXiv:1608.02662},
  2016.

\bibitem{SURFDocument}
J.~Heise, ``{The Sanford Underground Research Facility at Homestake},'' {\em J.
  Phys. Conf. Ser.}, vol.~606, no.~1, p.~012015, 2015.

\bibitem{Neves2017}
F.~Neves, A.~Lindote, A.~Morozov, {\em et~al.}, ``{Measurement of the absolute
  reflectance of polytetrafluoroethylene (PTFE) immersed in liquid xenon},''
  {\em J. Instrum.}, vol.~12, no.~01, p.~P01017, 2017.

\bibitem{LUXPositionReconstruction}
D.~Akerib {\em et~al.}, ``{Position Reconstruction in LUX},'' 2017.

\bibitem{FahamThesis}
C.~Faham, {\em {Prototype, Surface Commissioning and Photomultiplier Tube
  Characterization for the Large Underground Xenon (LUX) Direct Dark Matter
  Search Experiment}}.
\newblock PhD thesis, Brown University, 2013.

\bibitem{Faham_VUV_doublePE}
C.~Faham, V.~Gehman, A.~Currie, {\em et~al.}, ``{Measurements of
  wavelength-dependent double photoelectron emission from single photons in
  VUV-sensitive photomultiplier tubes},'' {\em J. Instrum.}, vol.~10, no.~09,
  p.~P09010, 2015.

\bibitem{LUXRun3Reanalysis}
D.~S. Akerib, H.~M. Ara\'ujo, X.~Bai, {\em et~al.}, ``{Improved Limits on
  Scattering of Weakly Interacting Massive Particles from Reanalysis of 2013
  LUX Data},'' {\em Phys. Rev. Lett.}, vol.~116, p.~161301, Apr 2016.

\bibitem{LUXKrPaper}
D.~S. Akerib, H.~M. Ara\'ujo, X.~Bai, {\em et~al.}, ``{$^{83m}$Kr Calibration
  of the 2013 LUX dark matter search},'' {\em Phys. Rev. D}, vol.~96, p.~112009,
  Dec 2017.

\bibitem{LUXDD}
D.~S. Akerib, H.~M. Ara{\'u}jo, X.~Bai, {\em et~al.}, ``{Low-energy (0.7{-}74
  keV) nuclear recoil calibration of the LUX dark matter experiment using D{-}D
  neutron scattering kinematics},'' {\em Submitted to Phys. Rev. C}.

\bibitem{LUXTritium}
D.~S. Akerib, H.~M. Ara\'ujo, X.~Bai, {\em et~al.}, ``Tritium calibration of
  the {LUX} dark matter experiment,'' {\em Phys. Rev. D}, vol.~93, p.~072009,
  Apr 2016.

\bibitem{AkashiRonquestPSD2015}
M.~Akashi-Ronquest, P.-A. Amaudruz, M.~Batygov, {\em et~al.}, ``{Improving
  photoelectron counting and particle identification in scintillation detectors
  with Bayesian techniques},'' {\em Astropart. Phys.}, vol.~65, pp.~40 -- 54,
  2015.

\bibitem{ROOT}
R.~Brun and F.~Rademakers, ``Root -- an object oriented data analysis
  framework,'' {\em Nucl. Instr. Meth. Phys. Res. A}, vol.~389, no.~1, pp.~81
  -- 86, 1997.

\bibitem{PMTHandbook}
Hamamatsu, {\em Photomultiplier Tubes: Basics and Applications}, 2007.

\bibitem{MongkolFlightPath}
M.~Moongweluwan, ``{The impact of photon flight path on S1 pulse shape analysis
  in liquid xenon two-phase detectors},'' {\em J. Instrum.}, vol.~11, no.~02,
  p.~C02036, 2016.

\bibitem{LUX_PRD}
D.~S. Akerib, H.~M. Ara\'ujo, X.~Bai, {\em et~al.}, ``Calibration, event
  reconstruction, data analysis and limits calculation for the {LUX} dark
  matter experiment,'' {\em arXiv:1712.05696}, 2017.

\bibitem{LUXSim_Paper}
D.~Akerib, X.~Bai, S.~Bedikian, {\em et~al.}, ``{LUXSim: A component-centric
  approach to low-background simulations },'' {\em Nucl. Instr. Meth. Phys.
  Res. A}, vol.~675, pp.~63 -- 77, 2012.

\bibitem{Geant4_1}
S.~Agostinelli, J.~Allison, K.~Amako, {\em et~al.}, ``Geant4--a simulation
  toolkit,'' {\em Nucl. Instr. Meth. Phys. Res. A}, vol.~506, no.~3, pp.~250 --
  303, 2003.

\bibitem{Geant4_2}
J.~Allison, K.~Amako, J.~Apostolakis, {\em et~al.}, ``Geant4 developments and
  applications,'' {\em IEEE Trans. Nucl. Sci.}, vol.~53, pp.~270--278, Feb
  2006.

\bibitem{R8778Datasheet}
Hamamatsu, {\em Photomultiplier Tube R8778}, 1 2008.
\newblock Rev. 3.

\bibitem{comsolRef}
COMSOL, Inc, {\em COMSOL Multiphysics Reference Manual, Version 5.3}.
\newblock {\texttt{http://www.comsol.com}}.

\bibitem{LenardoNEST}
B.~Lenardo, K.~Kazkaz, A.~Manalaysay, {\em et~al.}, ``{A Global Analysis of
  Light and Charge Yields in Liquid Xenon},'' {\em IEEE Trans. Nucl. Sci.},
  vol.~62, p.~3387, 2015.

\bibitem{MinuitPaper}
F.~James and M.~Roos, ``{Minuit: A System for Function Minimization and
  Analysis of the Parameter Errors and Correlations},'' {\em Comput. Phys.
  Commun.}, vol.~10, pp.~343--367, 1975.

\bibitem{lippincott2008}
W.~H. Lippincott, K.~J. Coakley, D.~Gastler, {\em et~al.}, ``Scintillation time
  dependence and pulse shape discrimination in liquid argon,'' {\em Phys. Rev.
  C}, vol.~78, p.~035801, Sep 2008.

\bibitem{ZEPLIN3_IDM}
D.~Akimov, H.~Ara{\'u}jo, E.~Barnes, {\em et~al.}, ``{{Limits on inelastic dark
  matter from ZEPLIN-III}},'' {\em Phys. Lett. B}, vol.~692, no.~3, pp.~180 --
  183, 2010.

\bibitem{DarksideFirstResult}
P.~Agnes, T.~Alexander, A.~Alton, {\em et~al.}, ``{{First results from the
  DarkSide-50 dark matter experiment at Laboratori Nazionali del Gran
  Sasso}},'' {\em Phys. Lett. B}, vol.~743, pp.~456 -- 466, 2015.

\end{thebibliography}
\end{document}